\documentclass[11pt,authoryear,review,3p,times]{elsarticle}
\usepackage[colorlinks=true,citecolor=blue,linkcolor=blue]{hyperref}
\usepackage[english]{babel}
\usepackage[utf8]{inputenc}
\usepackage[T1]{fontenc}
\usepackage{graphicx,array,bm,color,natbib,amssymb}

\newdefinition{definition}{Definition}
\newproof{proof}{Proof}
\usepackage{subcaption}
\journal{to the chosen journal}

\begin{document}
		
		\begin{frontmatter}
		\title{{ \LARGE Reliability of components of coherent systems: estimates in presence of masked data}}

		\author[label2]{\large Agatha S. Rodrigues\fnref{label3}\corref{aaa}}	
		\author{\large Carlos~Alberto~de~Bragan\c{c}a~Pereira\fnref{label2}}	
		\author{\large Adriano Polpo\fnref{label6}}
	
		\address[label2]{Institute of Mathematics and Statistics, University of S\~ao Paulo,  S\~ao Paulo, SP, Brazil.}
    	\address[label3]{Department of Obstetrics and Gynecology, S\~ao Paulo University Medical School, S\~ao Paulo, SP, Brazil.}	
		\address[label6]{Department of Statistics, Federal University of S\~ao Carlos, S\~ao Paulo, SP, Brazil.}

		\cortext[aaa]{e-mail: agatha@ime.usp.br\\ \indent\hspace{.15cm} phone number: +55 11 986667332}

\begin{abstract}
The reliability of a system of components depends on reliability of each component. Thus, the initial statistical work should be the estimation of the reliability of each component of the system. This is not an easy task because when the system fails, the failure time of a given component can not be observed, that is, censored data. \citet{AgathaPaper1} presented a solution for reliability estimation of components when it is avaliable the system failure time and the status of each component at the time of system failure (if it had failed before, after or it is responsible for system failure). However, there are situations it may be difficult to identify the status of components at the moment of system failure.
 Such cases are systems with masked causes of failure. Since parallel and series systems are the simplest systems, innumerous alternative solutions for these two systems have been appeared in the literature. To the best of our knowledge, this seems to be the first work that considers the general case of coherent systems. The three-parameter Weibull distribution is considered as the component failure time model. Identically distributed failure times is not required restrictions. Furthermore, there is no restriction on the subjective choice of prior distributions but preference has been given to continuous prior distributions; these priors represent well the nuances of the environment that the system operates. The statistical work of obtaining quantities of the posterior distribution is supported by the Metropolis within Gibbs algorithm. With several simulations, the excellent performance of the model was evaluated. We also consider a computer hard-drives real dataset in order to present the practical relevance of the proposed model.	

\end{abstract}
		
\begin{keyword}
	Component's reliability, Masked data, Coherent system, Bayesian three-parameter Weibull model, Metropolis within Gibbs algorithm.
\end{keyword}
\end{frontmatter}

\section{Introduction}
\label{secao1}


The first step in the study of the reliability of a system is the estimation of the reliability of each component of the system in order to obtain the highest system reliability. In general, the lifetime test could not be conducted on the components level, but in system level. Because of this, statistical inferences of component reliability is not an easy task. Considering a random sample of a system with $m$ components for which all $n$ sample units are observed up to death, every sample unit will produce a component failure time and a censored failure time for the remaining $m-1$ components, although the types of censoring could be different. For a specific component not responsible for one of the $n$ systems that failed at time $t$, either it is right-censored, in which case it still could continue working after $t$, or it is censored to the left if it has failed before $t$. Depending on system design (the way components are interconnected) and component's reliability, it is very common to have high percentages of censored data, sometimes greater than 80\%.

In the process of test, the available information are the $n$ system failure times and the status of each component at the time of systems failure (uncensored observation, right or left-censored). Approaches for components' reliability estimation for this situation have been proposed in literature. Some highlights are \cite{SalinasTorres}, \cite{PolpoCarlinhos}, \cite{CoqueJr}, \cite{PolpoSinhaSimoniCAB}, \cite{AgathaIC}, \cite{PolpoSiCar}, \cite{Bhering} and \cite{AgathaPaper1}. 

However, there are situations it may be difficult to identify the component that causes the system failure and, as a consequence, the status of the components at the moment of system failure. 
Cases like this are known as masked data failure cause and it is usually due to limited resources for the diagnosis of the cause of the failure. 
As a motivation, consider the reliability estimation of three components in computer hard-drive: eletronic hard, head flyability and disc magnetic \citep{Flehinger2002}. The first component to fail causes the failure of the system (computer hard-drive), that is, it is a series system with three components, and the analysis was performed in such a way that a small subset of components is identified as the possible cause of failure. In an attempt to repair the system as quickly as possible, the entire subset of components is replaced and the component responsible for the failure can not be identified.

For masked cause failure systems, \cite{Miyakawa} initially studied the reliability estimation for a two-component series system deriving the maximum likelhihood estimates (MLE) in closed form and a nonparametric estimates based on the Kaplan-Meier estimator.  \cite{UsherHodgson} extended Miyakawa' results to a three components in series. For series system and through MLE and Bayesian approaches under different parametric distributions, \cite{ReiserEtAl1995,GuessEtAl,LinUshGu1996,Sarhan2001,Sarhan2003,Usher1996} are good references. \cite{SarhanEl} estimated the reliability functions of the components that belong to a parallel system (the last component to fail causes the failure of the system), computing the maximum likelihood and Bayesian estimates. All these works assumed what one knowns as symmetry assumption, i.e., the probability of a system to be masked failure cause (masking probability) is the same regardless of the component that causes the system failure.

Under symmetry assumption relaxation, \cite{Muk2006} and \cite{Guttman1995} are good references. \cite{LinGuess1994} considered the masking probability in the likelihood construction and \cite{KuoYang2000} considered that, besides depending on the cause of system failure, these probabilities are decreasing functions on system failure time. 

The acelerated life tests (ALT) are usually used to obtain quickly information  about components' reliability. In such tests, components in a system are subjected to higher levels of stress to reduce its time to failure and use the result information to predict its behavior under normal conditions of operation. Considering ALT, \cite{Tan2009,HungHsu2012,Fan2014} proposed Bayesian estimators for masked series system. For hybrid systems (systems with components in series and in parallel), \cite{Wang2015} considered the MLE inference. 

\begin{figure}[h!]\centering
	\begin{minipage}[h]{0.3\linewidth}
		\includegraphics[width=0.6\linewidth]{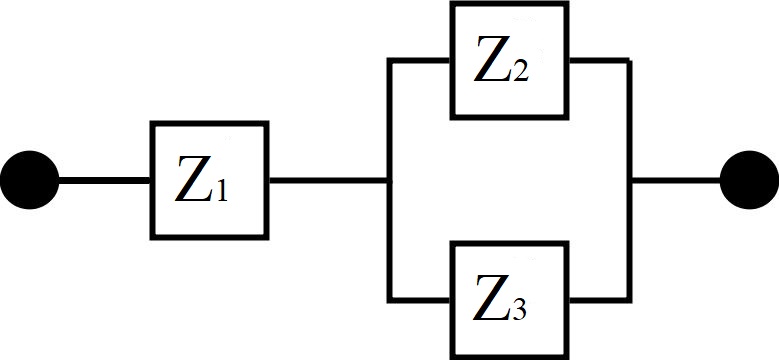}
		\subcaption{SPS representation} \label{sps}
	\end{minipage} 
	\begin{minipage}[h]{0.3\linewidth}
		\includegraphics[width=0.6\linewidth]{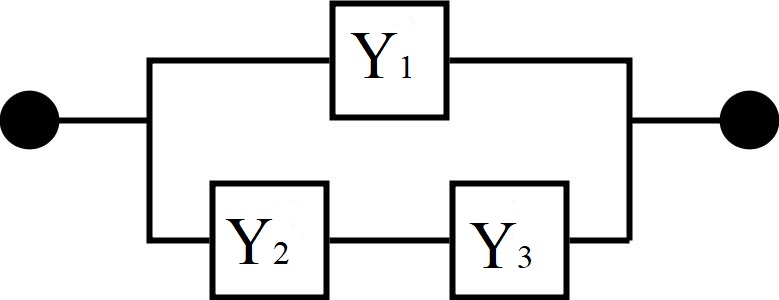}
		\subcaption{PSS representation} \label{pss}
	\end{minipage}
	\caption{SPS and PSS representations.}
	\label{systems_sps_pss}
\end{figure}

\begin{figure}[!ht]
	\centering
	\includegraphics[width=0.3\linewidth]{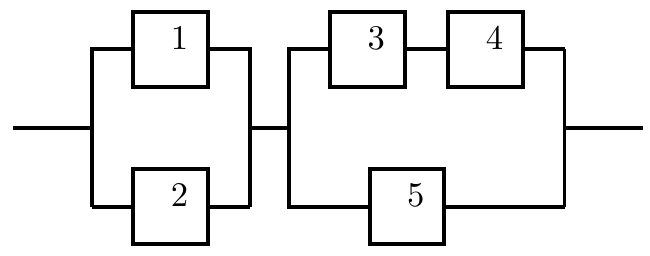}
	\caption{\label{chineses} Complex system with 5 components.}
\end{figure}

\begin{figure}[!ht]
	\centering
	\includegraphics[width=0.3\linewidth]{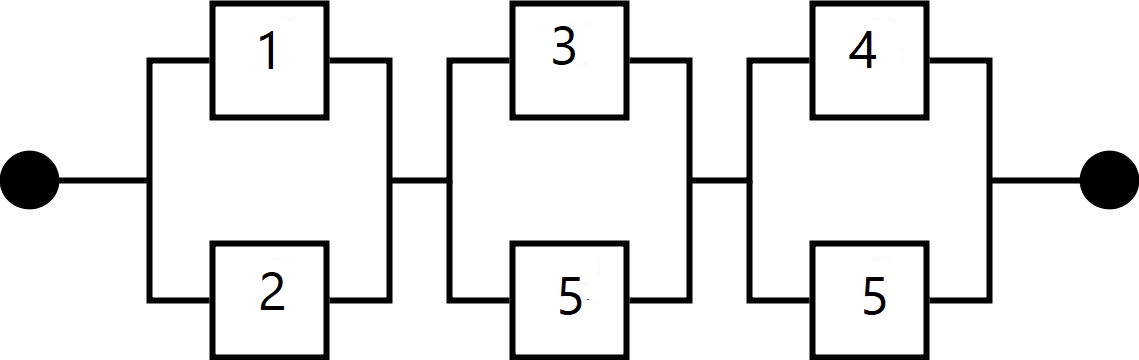}
	\caption{\label{chineses_SPS} SPS representation of system \ref{chineses}.}
\end{figure}

For components involved in series-parallel systems (SPS) and parallel-series systems (PSS) with three components, shown in Figure \ref{sps} and \ref{pss}, \citet{liu2017nonparametric} propose a Bayesian nonparametric estimators of the reliability functions with masked data under the ALT. They assume that components involved in SPS and PSS representations have mutually independent lifetimes and the proposed method can be considered for components from some complex coherent systems, once it is known that every coherent system can be writen as SPS and PSS representations \citep{BProschan}.
 The authors presents a discussion about estimation of components' reliability for the complex system in Figure \ref{chineses}, in which they estimate the reliabilities of components $j=1,2,5$ by representing the system as SPS representation (Figure \ref{sps}). Let $X_j$ be the lifetime of $j$th component ($j=1,\ldots,5$)  in Figure \ref{chineses} and $Z_l$ be the lifetime of $l$th component in SPS representation, $l=1,2,3$. Considering the component $j=5$, they build the simplified system by taking $Z_1=\max\{X_1,X_2\}$, $Z_2=\min\{X_3,X_4\}$ and $Z_3=X_5$. If the interest is the reliability estimation of components $j=3$ or $j=4$, the complex system can be represented as Figure \ref{chineses_SPS}. Let $Z_1=\min\{\max\{X_1,X_2\},\max\{X_4,X_5\}\}$ or $Z_1=\min\{\max\{X_1,X_2\},\max\{X_3,X_5\}\}$, $Z_2=X_3$ or $Z_2=X_4$ and $Z_3=X_5$; thus, the complex system can be simplified as the SPS in Figure \ref{sps}. However, their assumption of mutually independent components in the representation is violated because the presence of component $j=5$. Therefore, the method proposed by \citet{liu2017nonparametric} can not be considered for reliability estimation of components $j=3$ and $j=4$. 

\begin{figure}[!ht]
	\centering
	\includegraphics[width=0.3\linewidth]{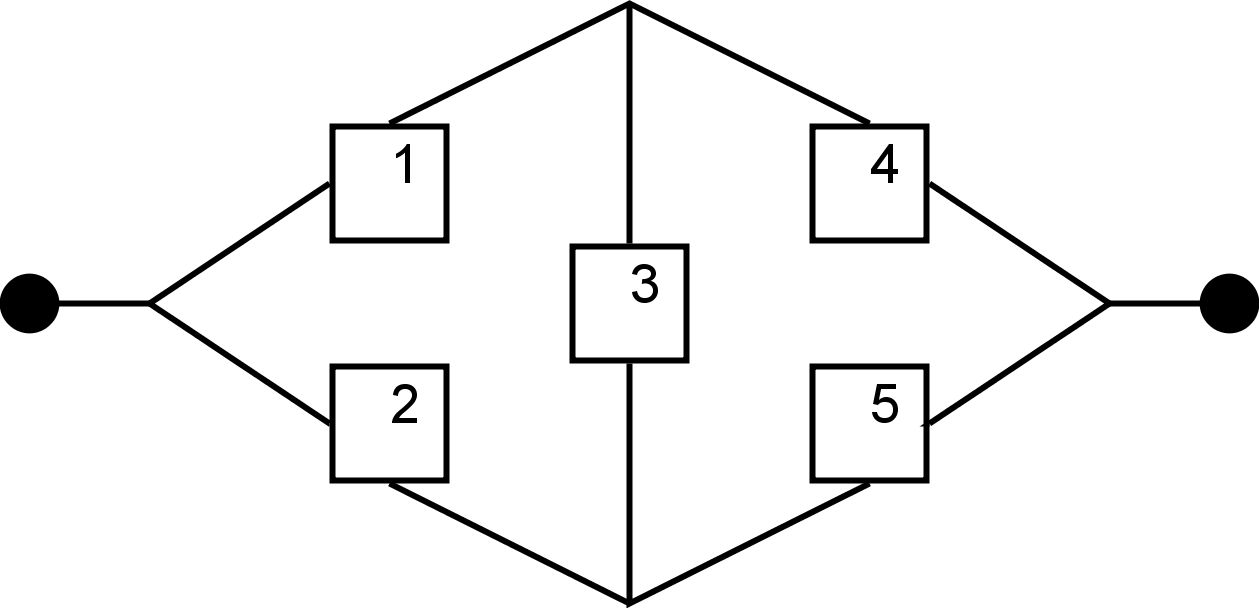}
	\caption{\label{fig:bridgenew} Bridge design.}
\end{figure}

\begin{figure}[h!]\centering
	\begin{minipage}[h]{0.4\linewidth}
		\includegraphics[width=0.7\linewidth]{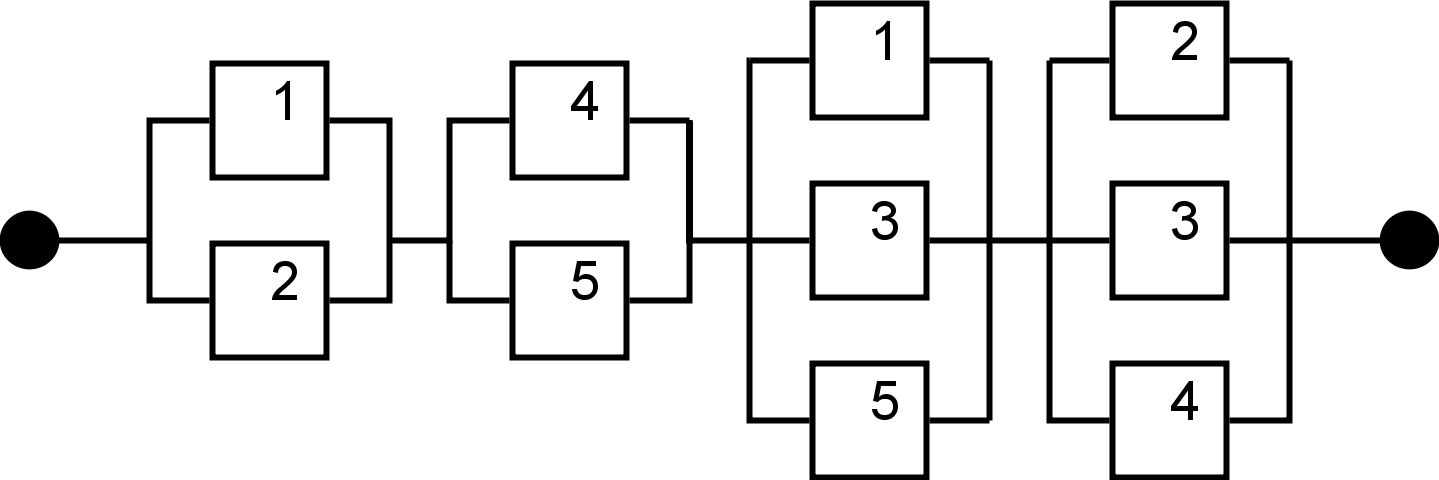}
		\subcaption{Bridge SPS representation} \label{bridge_SPS}
	\end{minipage} 
	\begin{minipage}[h]{0.4\linewidth}
		\includegraphics[width=0.5\linewidth]{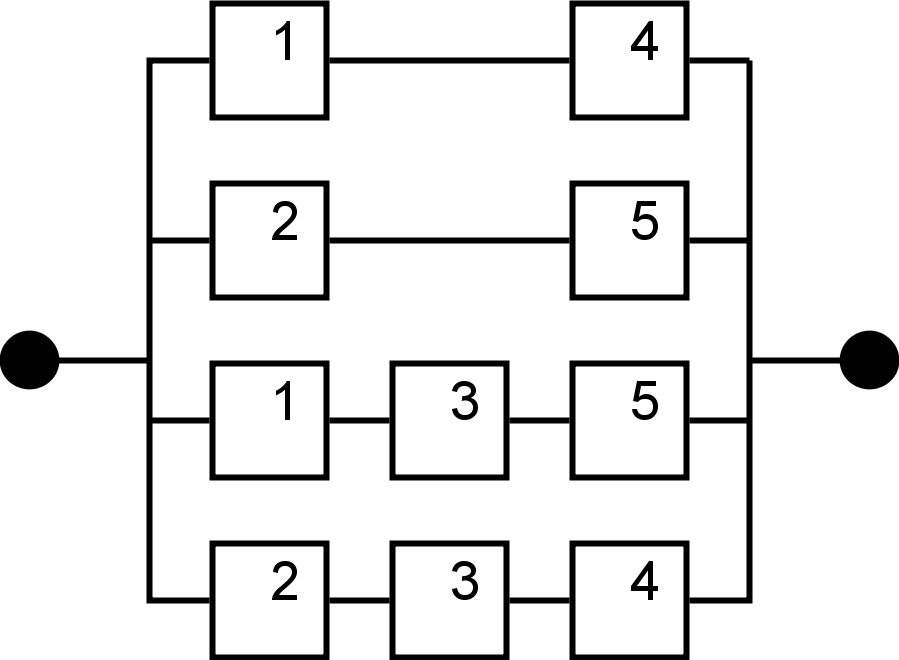}
		\subcaption{Bridge PSS representation} \label{bridge_PSS}
	\end{minipage}
	\caption{SPS and PSS representations.}
	\label{system_bridge_SPS_PSS}
\end{figure}

\begin{figure}[h!]\centering
	\begin{minipage}[h]{0.4\linewidth}
		\includegraphics[width=0.7\linewidth]{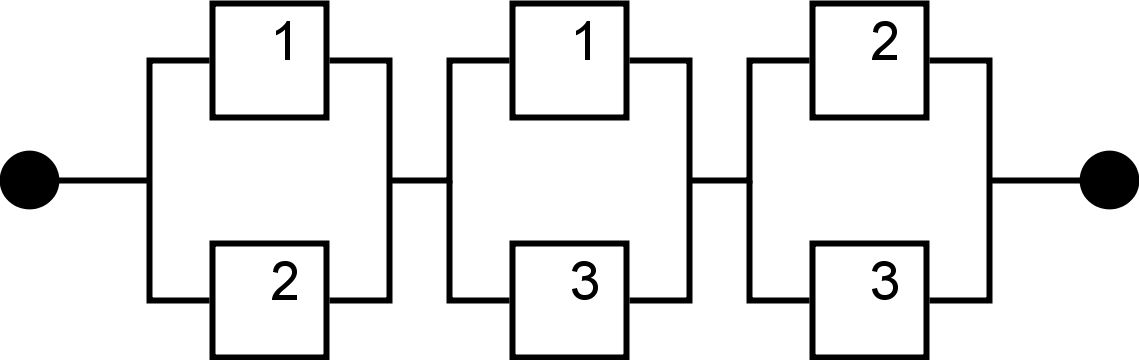}
		\subcaption{$2$-out-of-$3$ SPS representation} \label{2de3_SPS}
	\end{minipage} 
	\begin{minipage}[h]{0.4\linewidth}
		\includegraphics[width=0.5\linewidth]{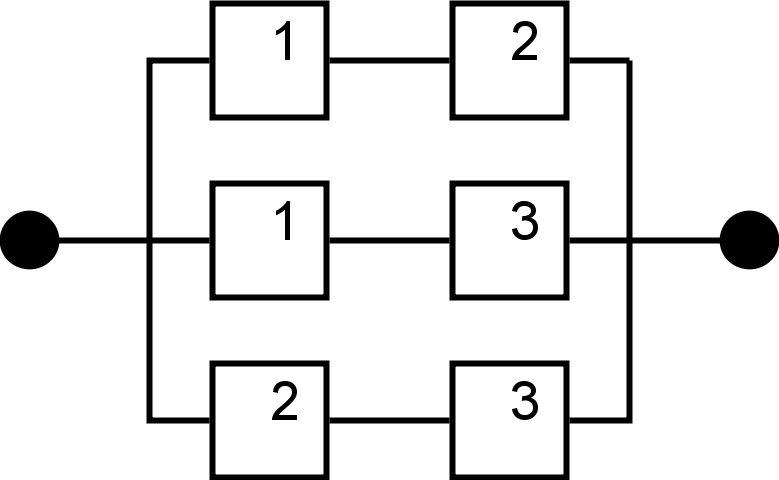}
		\subcaption{$2$-out-of-$3$ PSS representation} \label{2de3_PSS}
	\end{minipage}
	\caption{SPS and PSS representations of $2$-out-of-$3$.}
	\label{system_2de3_SPS_PSS}
\end{figure}

Others complex coherent systems present the same problem: some components may appear in two or more places in SPS or PSS representations. Figure \ref{fig:bridgenew} is the bridge system described in the literature \cite{BProschan} and  Figure \ref{system_bridge_SPS_PSS} illustrates its SPS and PSS representations. Note that each of the five  components appears twice for both representations. Another interesting design is the $k$-out-of-$m$ system (it works only if at least $k$ out of the $m$ components work). For instance, Figure \ref{system_2de3_SPS_PSS} considers the simple $2$-out-of-$3$ case into SPS and PSS representations. Note that each of the three components also appears twice in both combinations. Situations like these violate the assumption of \cite{liu2017nonparametric} and the estimator is not suitable for the reliability function of components involved in these complex coherent systems.


 The nonparametric estimator for reliability components involved in coherent system proposed by \cite{Sassa} can be considered in scenario with masked data, since the only necessary information is system failure time and system design, that is, it is not necessary to know the cause of failure, which is suitable for masked data situation. It happens because they assumed a restrictive assumption of components' lifetimes are {\it s}-independent and identically distributed and because of this, there is only one estimator for all different components, which can be a restrictive and not aplicable assumption.

To the best of our knowledge, we were not able to find works in literature that consider the reliability estimation of components involved in any coherent system with masked data in which identically distributed failure times is not imposed.
 In this sense, a Bayesian three-parameter Weibull model for component reliability in masked data is proposed. The presented model is general because can be considered for any coherent system, the symmetry assumption is not necessary and acelerated life tests (ALT) may also be considered. The statistical work of obtaining quantities of the posterior distribution is supported by the Metropolis within Gibbs algorithm. 
 
In the observed sample, the available information are systems failure times, system design and for some sample units, it is possible the diagnosis of the cause of the failure  and the status of components for these cases are observed. 
The performance of the component reliability estimator obtained from the proposed model is compared to the nonparametric estimator considered by \cite{Sassa} in scenarios of different proportion of masked data, complex system designs and different distribution for components lifetimes.
We also consider a real dataset in order to present the applicability of the proposed model. The dataset consists of $172$ computer hard-drives that were monitored over a period of $4$ years and their failure times were observed. However, for some of them ($38\%$) the cause of hard-drive fail was not identified. 

This paper is organized as follows. The proposed model and estimation method is described in Section \ref{secao2}. In Section \ref{secao_exemplos} we present simulated examples and a simulation study is presented in Section \ref{simulation_study}. In Section \ref{aplication} the applicability of the proposed model is presented in computer hard-drives problem. Finally, some final remarks and additional comments are given in Section \ref{final}.

\section{Weibull Model and Estimation Method}\label{secao2}

Consider a system with $m$ components and let the index $j$ representing the $j$th component. The failure time of each component can be censored or not. Let $X_j$ be a random variable for the failure time of the $j$-th component, $t$ the failure time of the system, and $\delta_j$ an indicator of censor. We assume that $X_1,X_2,\ldots,X_m$ are mutually independent. The observation of $X_j$ can be: $X_j = t$, the failure time of $X_j$ it is not censored ($\delta_j = 1$); $X_j > t$, the failure time is right censored ($\delta_j = 2$); and $X_j \leq t$, the failure time is left censored ($\delta_j = 3$). Also, $j$-th component can belong to the masked set or not.

Let $\Upsilon$ be the set of index indicating possible components that produced the failure of the system, that is, components that have their failure time masked, $\Upsilon$ is a subset of $\{1, \ldots, m\}$. Let $t_1, \ldots, t_n$ a sample of system failure time of size $n$, and $\Upsilon_i$ is the set of masked components in the $i$-th sample, $i = 1, \ldots, n$. Also, $\upsilon_{ji} = 1$ if the $j$-th component has the failure time masked ($j \in \Upsilon_i$), and $\upsilon_{ji} = 0$ otherwise ($j \notin \Upsilon_i$), $j = 1, \ldots, m$. The observation of $j$-th component will be one of the following:

\begin{description}
	\item[uncensored; not masked:] $\delta_{ji} = 1$ and $\upsilon_{ji} = 0$;
	\item[right censored; not masked:] $\delta_{ji} = 2$ and $\upsilon_{ji} = 0$;
	\item[left censored; not masked:] $\delta_{ji} = 3$ and $\upsilon_{ji} = 0$;
	\item[masked:]  and $\upsilon_{ji} = 1$.
\end{description}
If a component has the failure time masked, the component can be the one that produced the failure of the system (uncensored), right censored or left censored. Consider that
\begin{eqnarray}
{\lambda_1}_j(t) &= \Pr(\upsilon_j = 1 \mid t, \delta_j = 1), \nonumber \\ 
{\lambda_2}_j(t) &= \Pr(\upsilon_j = 1 \mid t, \delta_j = 2),  \nonumber \\ 
{\lambda_3}_j(t) &= \Pr(\upsilon_j = 1 \mid t, \delta_j = 3), \nonumber
\end{eqnarray}
where ${\lambda_1}_j(t)$ is the conditional probability of the $j$-th component be masked, given the failure time of the system $t$, and the censor type $\delta_j = 1$. ${\lambda_2}_j(t)$ and ${\lambda_3}_j(t)$ are analogous to ${\lambda_1}_j(t)$. Here, we consider that ${\lambda_1}_j(t) = {\lambda_1}_j$, ${\lambda_2}_j(t) = {\lambda_2}_j$, and ${\lambda_3}_j(t) = {\lambda_3}_j$, that is, the probability of a component be masked does not depend on the failure time $t$.

For each component, we can observe the triple $(t_i, \delta_{ji}, \upsilon_{ji})$, $i, \ldots, n$. Our interest consist in the estimation of the distribution function, $F$, of the $j$-th component, $X_j$. We consider a parametric family model for $F$ with parameter $\bm{\theta}_j$ (scalar or vector), then the estimation of the parameter $\bm{\theta}_j$ induces the distribution function $F$. The available information from the data is one of the following types:
\begin{enumerate}
	\item $\Pr(X_{ij} \in (t_i, t_i], \upsilon_i = 0 \mid \bm{\theta}_j) = f(t_i \mid \bm{\theta}_j) (1- {\lambda_1}_j)$, if the $i$-th observation is uncensored and not masked;
	\item $\Pr(X_{ij} \in (t_i, \infty), \upsilon_i = 0 \mid \bm{\theta}_j) = [1-F(t_i \mid \bm{\theta}_j)](1- {\lambda_2}_j)$, if the $i$-th observation is right censored and not masked;
	\item $\Pr(X_{ij} \in (0, t_i], \upsilon_i = 0 \mid \bm{\theta}_j) = F(t_i \mid \bm{\theta}_j) (1- {\lambda_3}_j)$, if the $i$-th observation is left censored and not masked;
	\item $\Pr(X_{ij} \in (t_i, t_i], \upsilon_i = 1 \mid \bm{\theta}_j) = f(t_i \mid \bm{\theta}_j) {\lambda_1}_j$, if the $i$-th observation is uncensored and masked;
	\item $\Pr(X_{ij} \in (t_i, \infty), \upsilon_i = 1 \mid \bm{\theta}_j) = [1-F(t_i \mid \bm{\theta}_j)] {\lambda_2}_j$, if the $i$-th observation is right censored and masked; and
	\item $\Pr(X_{ij} \in (0, t_i], \upsilon_i = 1 \mid \bm{\theta}_j) = F(t_i \mid \bm{\theta}_j) {\lambda_3}_j$, if the $i$-th observation is left censored and masked.
\end{enumerate}
However, we do not have information about the cases 4 to 6, since when the data is masked, we do not know if that component was censored or not. 
Consider an augmented data procedure (latent variable), define $d_{1ji} = 1$ if the masked observation is not censored or $d_{1ji} = 0$ otherwise, $d_{2ji} = 1$ if the masked observation is right censored or $d_{2ji} = 0$ otherwise, and $d_{3ji} = 1$ if the masked observation is left censored or $d_{3ji} = 0$ otherwise. Besides, ${\bm d}_{ji}=(d_{1ji},d_{2ji},d_{3ji})$ and $\sum_{l=1}^3d_{lji}=1$.

Let $R(t_i \mid \bm{\theta}_j) = 1-F(t_i \mid \bm{\theta}_j)$ the reliability function. The likelihood function of the $j$-th component can be writen as a part of non-masked data and a part for masked data (augmented data), 
\begin{eqnarray}
& L(\bm{\theta}_j, {\lambda_1}_j, {\lambda_2}_j, {\lambda_3}_j, \bm{d}_j \mid \bm{t}, \bm{\delta}_j, \bm{\upsilon}_j)  = \prod\limits_{i: ~ \upsilon_{ji} = 0} \Big\{ \big[f(t_i \mid \bm{\theta}_j) ~ (1- {\lambda_1}_j)\big]^{I(\delta_{ji} = 1)} \big[R(t_i \mid \bm{\theta}_j) ~ (1- {\lambda_2}_j)\big]^{I(\delta_{ji} = 2)} \nonumber \\
& \times \big[F(t_i \mid \bm{\theta}_j) ~ (1- {\lambda_3}_j)\big]^{I(\delta_{ji} = 3)} \Big\} \prod\limits_{i: ~ \upsilon_{ji} = 1} \Big\{ \big[f(t_i \mid \bm{\theta}_j) ~ {\lambda_1}_j \big]^{d_{1ji}}  \big[R(t_i \mid \bm{\theta}_j) ~ {\lambda_2}_j \big]^{d_{2ji}}  \big[F(t_i \mid \bm{\theta}_j) ~ {\lambda_3}_j \big]^{d_{3ji}} \Big\}, \label{vero_masked}
\end{eqnarray}
where $I(A) = 1$ if $A$ is true and $0$ otherwise, $\bm{t} =$ $\{t_1, \ldots, t_n\}$, $\bm{\upsilon}_j =$ $\{\upsilon_{j1}, \ldots, \upsilon_{jn}\}$, $\bm{d}_j = ({\bm d}_{ji}:i \in \{\upsilon_{ji} = 1\})$ and $\bm{\delta}_j = (\delta_{ji}:i \in \{\upsilon_{ji} = 0\})$.

The likelihood function in (\ref{vero_masked}) is generic and straightforward for any probability distribution. The distribution considered is the three-parameter Weibull. The Weibull distribution has characteristics that make this distribution a great candidate to model components lifetimes. One of them is that by changing parameter values the distribution takes a variety of shapes and it has important distributions as special cases, besides allowing modeling increasing, decreasing or constant hazard rates \cite{Rinne}. 

The reliability function is as follows:
    \begin{eqnarray}
    	R(t \mid \bm{\theta_j}) = \exp\left[-\left(\frac{t-\mu_j}{\eta_j}\right)^{\beta_j}\right], \nonumber
    \end{eqnarray}
   for $t > 0$, where $\bm{\theta_j}= (\beta_j, \eta_j, \mu_j)$ and $\beta_j > 0$ (shape), $\eta_j > 0$ (scale) and $0<\mu_j<t$ (location). 

	The Weibull distribution with two parameters ($\mu_j=0$) is the most celebrated case in the literature. However, the location parameter that represents the baseline lifetime has an important meaning in reliability and survival analysis. In reliability, a component under test may not be new. In medicine, for instance, a patient may have the disease before the onset medical appointment. Not taking account the initial time can underestimate the other parameters. Clearly, for a new component testing $\mu_j$ may be 0.

The posterior distribution of $(\bm{\theta}_j, {\lambda_1}_j, {\lambda_2}_j, {\lambda_3}_j, \bm{d}_j)$ comes out to be
\begin{eqnarray} 
	\pi(\bm{\theta}_j, {\lambda_1}_j, {\lambda_2}_j, {\lambda_3}_j, \bm{d}_j  \mid \bm{t},\bm{\delta}_j, \bm{\upsilon}_j) & ~~ \propto &  ~~ \pi(\bm{\theta}_j, {\lambda_1}_j, {\lambda_2}_j, {\lambda_3}_j, \bm{d}_j)  L(\bm{\theta}_j, {\lambda_1}_j, {\lambda_2}_j, {\lambda_3}_j, \bm{d}_j \mid \bm{t}, \bm{\delta}_j, \bm{\upsilon}_j),    \label{posteriori}
\end{eqnarray}
where $\pi(\bm{\theta}_j, {\lambda_1}_j, {\lambda_2}_j, {\lambda_3}_j, \bm{d}_j)$ is the prior distribution of $(\bm{\theta}_j, {\lambda_1}_j, {\lambda_2}_j, {\lambda_3}_j, \bm{d}_j)$. The prior distributions of all parameters are considered independent with gamma distribution with mean $1$ and variance $1000$ for $\beta_j$, $\eta_j$, $\mu_j$ and uniform distribution over $(0,1)$ for ${\lambda_1}_j$, ${\lambda_2}_j$ and ${\lambda_3}_j$. Besides, $\Pr(d_{lji}=1)=\Pr(d_{lji}=0)=0.5$, for $l=1,2,3$.

 In this paper, no prior information about component's operation is known and noninformative prior is considered. However, it is possible to express a prior information about the component functioning in the system through the opinion of an expert and/or through past experiences. 

The posterior density in Equation (\ref{posteriori}) has not close form. An alternative is to rely on Markov chain Monte Carlo (MCMC) simulations. Here we consider Metropolis within Gibbs algorithm. This algorithm is suitable in this situation because it is possible direct sampling from conditional distribution for some parameters but for others this is not possible \citep{Tierney}. The algorithm works in the following steps:

\vspace{0.2cm}
{\it
	\begin{enumerate}
		\item Attribute initial values $\bm{\theta}_j^{(0)}$, $\lambda_{1j}^{(0)}$, $\lambda_{2j}^{(0)}$ and $\lambda_{3j}^{(0)}$ for   $\bm{\theta}_j=(\beta_j,\eta_j,\mu_j)$, $\lambda_{1j}$, $\lambda_{2j}$ and $\lambda_{3j}$, respectively, and set $b=1$;
		\item For $i \in \bm{\upsilon}_j$, draw $\bm{d}_{ji}^{(b)}$ from $\pi(\bm{d}_{ji} \mid \bm{t}, \bm{\delta}_j, \bm{\upsilon}_j, \bm{\theta}_j^{(b-1)},\lambda_{1j}^{(b-1)},\lambda_{2j}^{(b-1)},\lambda_{3j}^{(b-1)})$, in which:
		\begin{eqnarray} 
			\pi(\bm{d}_{ji} \mid \bm{t}, \bm{\delta}_j, \bm{\upsilon}_j, \bm{\theta}_j,\lambda_{1j} ,\lambda_{2j},\lambda_{3j} ) & ~~ \propto &  ~~   \Big\{\lambda_{1j} f(t_{i}|\bm{\theta}_j) \Big\}^{d_{1ji}}
			\Big\{\lambda_{2j}R(t_{i}|\bm{\theta}_j)\Big\}^{d_{2ji}}\Big\{\lambda_{3j}F(t_{i}|\bm{\theta}_j)\Big\}^{d_{3ji}}, \nonumber                 
		\end{eqnarray}
		that is, $\bm{d}_{ji} \mid \bm{t},\bm{\delta}_j,\bm{\upsilon}_j,\bm{\theta}_j,\lambda_{1j} ,\lambda_{2j},\lambda_{3j} \sim \mbox{Multinomial}(1;p_{1ji},p_{2ji},p_{3ji})$ in which $p_{1ji}=\lambda_{1j}f(t_{i}|\bm{\theta}_j)/C$, $p_{2ji}=\lambda_{2j} R(t_{i}|\bm{\theta}_j)/C$ and $p_{3ji}=\lambda_{3j}F(t_{i}|\bm{\theta}_j)/C$, where $C=\lambda_{1j}f(t_{i}|\bm{\theta}_j)+\lambda_{2j} R(t_{i}|\bm{\theta}_j)+\lambda_{3j}F(t_{i}|\bm{\theta}_j)$;
		\item Draw $\bm{\theta}_j^{(b)}$ from $\pi(\bm{\theta}_j\mid \bm{t}, \bm{\delta}_j, \bm{\upsilon}_j,\bm{d}_j^{(b)},\lambda_{1j}^{(b-1)} ,\lambda_{2j}^{(b-1)} ,\lambda_{3j}^{(b-1)})$ through Metropolis-Hastings algorithm \citep{RobertCasella}, where 
		\begin{eqnarray} 
			&& \pi(\bm{\theta}_j\mid \bm{t}, \bm{\delta}_j, \bm{\upsilon}_j,\bm{d}_j,\lambda_{1j} ,\lambda_{2j} ,\lambda_{3j})    \propto  ~~ \pi(\bm{\theta}_j) \prod\limits_{i: ~ \upsilon_{ji} = 0} \Big\{ \big[f(t_i \mid \bm{\theta}_j) ~ (1- {\lambda_1}_j)\big]^{I(\delta_{ji} = 1)} \big[R(t_i \mid \bm{\theta}_j) ~ (1- {\lambda_2}_j)\big]^{I(\delta_{ji} = 2)} \nonumber \\
			&\times& \big[F(t_i \mid \bm{\theta}_j) ~ (1- {\lambda_3}_j)\big]^{I(\delta_{ji} = 3)} \Big\} \prod\limits_{i: ~ \upsilon_{ji} = 1} \Big\{ \big[f(t_i \mid \bm{\theta}_j) ~ {\lambda_1}_j \big]^{d_{1ji}}  \big[R(t_i \mid \bm{\theta}_j) ~ {\lambda_2}_j \big]^{d_{2ji}}  \big[F(t_i \mid \bm{\theta}_j) ~ {\lambda_3}_j \big]^{d_{3ji}} \Big\}.     \nonumber
		\end{eqnarray}
		
		\item Simulate $\lambda_{1j}$ from $\pi(\lambda_{1j}\mid  \bm{t}, \bm{\delta}_j, \bm{\upsilon}_j,\bm{d}_j^{(b)},\bm{\theta}_j^{(b)} ,\lambda_{2j}^{(b-1)} ,\lambda_{3j}^{(b-1)})$ in which
		\begin{eqnarray} 
			\pi(\lambda_{1j}\mid \bm{t}, \bm{\delta}_j, \bm{\upsilon}_j,\bm{d}_j,\bm{\theta}_j ,\lambda_{2j},\lambda_{3j}) & ~~ \propto &  ~~   \lambda_{1j}^{\sum\limits_{i: ~ \upsilon_{ji} = 1}d_{1ji}}(1-\lambda_{1j})^{n_f},    \nonumber                 
		\end{eqnarray}
		that is, $\lambda_{1j}\mid (\bm{t}, \bm{\delta}_j, \bm{\upsilon}_j,\bm{d}_j,\bm{\theta}_j ,\lambda_{2j} ,\lambda_{3j}) \sim \mbox{Beta}(\sum\limits_{i: ~ \upsilon_{ji} = 1}d_{1ji}+1,n_f+1)$, in which $n_f$ is the number of systems in which component $j$ is the known to responsible of system failure.

		\item Simulate $\lambda_{2j}$ from $\pi(\lambda_{2j}\mid  \bm{t}, \bm{\delta}_j, \bm{\upsilon}_j,\bm{d}_j^{(b)},\bm{\theta}_j^{(b)} ,\lambda_{1j}^{(b-1)} ,\lambda_{3j}^{(b-1)})$ in which
		\begin{eqnarray} 
		\pi(\lambda_{2j}\mid \bm{t}, \bm{\delta}_j, \bm{\upsilon}_j,\bm{d}_j,\bm{\theta}_j ,\lambda_{1j},\lambda_{3j}) & ~~ \propto &  ~~   \lambda_{2j}^{\sum\limits_{i: ~ \upsilon_{ji} = 1}d_{2ji}}(1-\lambda_{2j})^{n_r},    \nonumber                 
		\end{eqnarray}
		that is, $\lambda_{2j}\mid (\bm{t}, \bm{\delta}_j, \bm{\upsilon}_j,\bm{d}_j,\bm{\theta}_j ,\lambda_{1j} ,\lambda_{3j}) \sim \mbox{Beta}(\sum\limits_{i: ~ \upsilon_{ji} = 1}d_{2ji}+1,n_r+1)$, in which $n_r$ is the number of systems in which component $j$ is observed to be right-censored.		
		\item Simulate $\lambda_{3j}$ from $\pi(\lambda_{3j}\mid  \bm{t}, \bm{\delta}_j, \bm{\upsilon}_j,\bm{d}_j^{(b)},\bm{\theta}_j^{(b)} ,\lambda_{1j}^{(b-1)} ,\lambda_{2j}^{(b-1)})$ in which
		\begin{eqnarray} 
		\pi(\lambda_{3j}\mid \bm{t}, \bm{\delta}_j, \bm{\upsilon}_j,\bm{d}_j,\bm{\theta}_j ,\lambda_{1j},\lambda_{2j}) & ~~ \propto &  ~~   \lambda_{3j}^{\sum\limits_{i: ~ \upsilon_{ji} = 1}d_{3ji}}(1-\lambda_{3j})^{n_f},    \nonumber                 
		\end{eqnarray}
		that is, $\lambda_{3j}\mid (\bm{t}, \bm{\delta}_j, \bm{\upsilon}_j,\bm{d}_j,\bm{\theta}_j ,\lambda_{1j} ,\lambda_{2j}) \sim \mbox{Beta}(\sum\limits_{i: ~ \upsilon_{ji} = 1}d_{3ji}+1,n_f+1)$, in which $n_f$ is the number of systems in which component $j$ is observed to be left-censored.
		\item Let $b=b+1$ and repeat steps $2)$ to $7)$ until $b=B$, where $B$ is pre-set number of simulated samples of $(\bm{\theta}_j,\lambda_{1j},\lambda_{2j},\lambda_{3j},\bm{d}_j)$.
	\end{enumerate}
}

\vspace{0.2cm}

When discarding burn-in sample (first generated values discarded to eliminate the effect of the assigned initial values for parameters) and jump sample (spacing among generated values to avoid correlation problems), a sample of size $n_p$ from the joint posterior distribution of $(\bm{\theta}_j,\lambda_{1j},\lambda_{2j},\lambda_{3j},\bm{d}_j)$ is obtained. For the $j$th component, the sample from the posterior can be expressed as $(\bm{\theta}_{j1},\bm{\theta}_{j2},\ldots,\bm{\theta}_{jn_p})$, $(\lambda_{1j1},\lambda_{1j2},\ldots,\lambda_{1jn_p})$, $(\lambda_{2j1},\lambda_{2j2},\ldots,\lambda_{2jn_p})$ and $(\lambda_{3j1},\lambda_{3j2},\ldots,\lambda_{3jn_p})$ and thus, posterior quantities of reliability function $R(t\mid\bm{\theta}_{j})$ of interest can be easily obtained \citep{RobertCasella}. For example, the posterior mean is given by
\begin{eqnarray}
	{\rm E}[R(t\mid \bm{\theta}_{j}) \mid Data] = \frac{1}{n_p}\sum_{k=1}^{n_p}{R(t \mid \bm{\theta}_{jk})},~~ \mbox{for each} ~ t > 0. \label{relia_bayes}
\end{eqnarray}

\subsection{Symmetric Masking Probabilities}
The assumption that the masking probabilities are the same regardless of cause of failure, that is, $\lambda_{1j}=\lambda_{2j}=\lambda_{3j}$ is plausible for some masked data system situations.

Under this assumption, we have that $\bm{d}_{ji} \mid \bm{t},\bm{\delta}_j,\bm{\upsilon}_j,\bm{\theta}_j,\lambda_{1j} ,\lambda_{2j},\lambda_{3j} \sim \mbox{Multinomial}(1;p_{1ji},p_{2ji},p_{3ji})$ in which $p_{1ji}=f(t_{i}|\bm{\theta}_j)/C$, $p_{2ji}=R(t_{i}|\bm{\theta}_j)/C$ and $p_{3ji}=F(t_{i}|\bm{\theta}_j)/C$, where $C=1+f(t_{i}|\bm{\theta}_j)$. That is, the estimation process does not depend on masking probabilities $\lambda_{lj}$, $l=1,2,3$, anymore and the algorithm presented previously can be considered eliminating steps $4)$ to $6)$. 

\subsection{Incorporation of Covariates}

Imagine that in a sample of $n$ systems, their units are not exposed  exactly to the same temperature and pressure conditions, for example, and, depending on their values, it can increase or decrease the reliability of the components. Thus, it is important to take into account these different conditions in the reliability estimation of each component and this is possible by incorporating covariates in the model.
	
In general, evaluating the performance of components in a system under normal conditions of use can be time-consuming and costly. For this reason, another importance emerges from the incorporation of covariables: accelerated life tests in which covariables are called stress variables.

In accelerated life tests (ALT), the components are subjected to stress levels sufficient to reduce their time to failure and inferences are obtained about their behavior under normal operating conditions. 

The analysis of stress-response relationships and extrapolation to usual operating conditions can be done through regression models for data from accelerated tests, called accelerated life models.
In ALT models, we have multiplicative effect with reliability time, that is, $R(t)=R_0(\varphi t)$, where $\varphi$ is the acceleration factor and $R_0(t)$ is the baseline reliability function. Thus, if $\varphi>1$, $R(t)$ behaves as $R_0(t)$ ``in the future'', if $\varphi<1$, $R(t)$ behaves as $R_0(t)$ ``in the past'' and if $\varphi=1$, $R(t)=R_0(t)$.

Some parametric models have the property of accelerated life test. The three-parameter Weibull distribution is a ALT model, in which $R_j(t\mid \bm{\theta_j})=R_0((t-\mu_j)\varphi_j\mid \beta_j)$ and $R_0(\cdot\mid \beta_j)$ is the reliability function of a Weibull distribution with scale $1$ and shape $\beta_j$.

We consider the inclusion of covariates in the scale parameter through a log link function, that is, $\eta_{j}=\exp(\bm{w}_{j}^\top\bm{\gamma}_j)$, in which $\bm{\gamma}_j$ is a vector $k\times 1$ of regression coefficients and $\bm{w}_{j}$ is a vector of covariates for $j$-th component. 

\section{Simulated System Datasets} \label{secao_exemplos}

We consider three simulated examples of complex system structure presented in Figures \ref{chineses}, \ref{fig:bridgenew} and \ref{system_2de3_SPS_PSS}. As mentioned previously, there is no solution in the literature for reliability estimation of all components involved in  these complex systems. 

Once the design of the systems is observed, $\Upsilon_i$ consists of components that no longer work in the system failure, that is, there is no right-censored for components in $\Upsilon_i$, which leads to  $\lambda_{2j}=0$. Besides, $j \in \Upsilon_i$ only if $j$ belongs to the minimal cut that caused the $i$-th system fail. A cut set is a set of components which by failing causes the system to fail. A cut set is said to be a minimal if it can not be reduced without losing its status as a cut set. For the bridge system represented in Figure \ref{fig:bridgenew}, for example, we have four minimal cut set, they are: $\{1,2\}$, $\{4,5\}$, $\{1,3,5\}$ and $\{2,3,4\}$. 

Imagine a situation that this bridge system fail and the components 1, 2 and 3 do not work at the moment of system failure. Then, only components 1 and 2 belong to set $\Upsilon$, once the component 3 does not belong to the minimal cut that caused the system failure and, in fact, the component 3 is observed to be left-censored failure time. We fitted the proposed model under symmetric assumption (that is, $\lambda_{1j}=\lambda_{3j}$) and no covariates in the model.


The three simulated systems have the following characteristics:
\begin{itemize}
	\item System structure 1 ($2$-out-of-$3$): $m=3$ and $X_{1}$ generated from Weibull distribution with mean $15$ and variance $8$, $X_{2}$ from gamma distribution with mean $18$ and variance $12$, $X_{3}$ from lognormal distribution with mean $20$ and variance $10$ and the system failure time is $\break$ $T=\max \{\min\{X_{1},X_{2}\}, \min\{X_{1},X_{3}\}, \min\{X_{2},X_{3}\}\}$. Besides, $n=300$ and the proportion of masked system is $p=0.4$. 
	\item System structure 2: $m=5$ and $X_{1}$ generated from Weibull distribution with mean $12$ and variance $15$, $X_{2}$ from gamma distribution with mean $11$ and variance $11$, $X_{3}$ from three-parameter Weibull distribution with mean $12$ and variance $9$, $X_{4}$ from lognormal distribution with mean $12$ and variance $7$ and $X_{5}$ from three-parameter Weibull distribution with mean $11$ and variance $14$. In this structure, the system lifetime is given by $T = \min \{\max\{X_{1},X_{2}\}, \max\{\min\{X_{3},X_{4}\},X_{5}\}\}$. For this case, $n=100$ and the proportion of masked data systems is $p=0.3$.
	\item System structure 3 (bridge system): $m=5$ and $X_{1}$ generated from Weibull distribution with mean $4$ and variance $15$, $X_{2}$ from a modified Weibull distribution \citep{WeibMod} with mean $5.6$ and variance $14.9$, $X_{3}$ from lognormal distribution with mean $6$ and variance $7$, $X_{4}$ from gamma distribution with mean $5$ and variance $8$ and $X_{5}$ from three-parameter Weibull distribution with mean $4$ and variance $8$. In this structure, the system lifetime is given by $\break$ $T = \max \{\min\{X_{1},X_{4}\}, \min\{X_{2},X_{5}\},\min\{X_{1},X_{3},X_{5}\},\min\{X_{2},X_{3},X_{4}\}\}$. Besides, $n=50$ and the proportion of masked data systems is $p=0.2$.
\end{itemize}

%
%

For all systems, we generated $30000$ values of each parameter, disregarding the first $10000$ iterations to eliminate the effect of the initial values and spacing of size $20$ to avoid correlation problems, obtaining a sample of size $n_p=1000$. The chains convergence was monitored and good convergence results were obtained.

The proposed model is compared to the nonparametric  approach proposed by \cite{Sassa}. This method estimates the reliability of components involved in any coherent system, from the simplest to the most complex. The only necessary information is system design and system failure time, that is, it is not necessary to know the cause of failure which is suitable for masked data situation. It happens because they assumed a restrictive assumption of components' lifetimes are {\it s}-independent and identically distributed and because of this, there is only one estimate for all different components. For simplification we refer to this estimator as BSNP.

We evaluate the mean absolute error (MAE) from the estimators to the true distribution as the comparison measure.  $R(t)$ and $\widehat{R}(t)$ are the true reliability function and its estimate, respectively. Hence the MAE is evaluated by $\frac{1}{l}\sum_{\ell=1}^{l} \mid \widehat{R}(g_{\ell})-R(g_{\ell}) \mid$, where $\{g_1, \ldots, g_{\ell}, \ldots, g_l \}$ is a grid in the space of failure times.


\subsection{System 1} \label{ex_simu_2de3_texto}


The posterior quantities of $R(t|\mbox{\boldmath{$\theta_j$}})$ for some values of $t$ are shown in Table \ref{medidas_posterior_2de3}. In the following of this paper the posterior mean (W3PM) is considered as the performing posterior measure of reliability function obtained by the proposed model.

\begin{figure}[h!]\centering
	\begin{minipage}[b]{0.32\linewidth}
		\includegraphics[width=\linewidth]{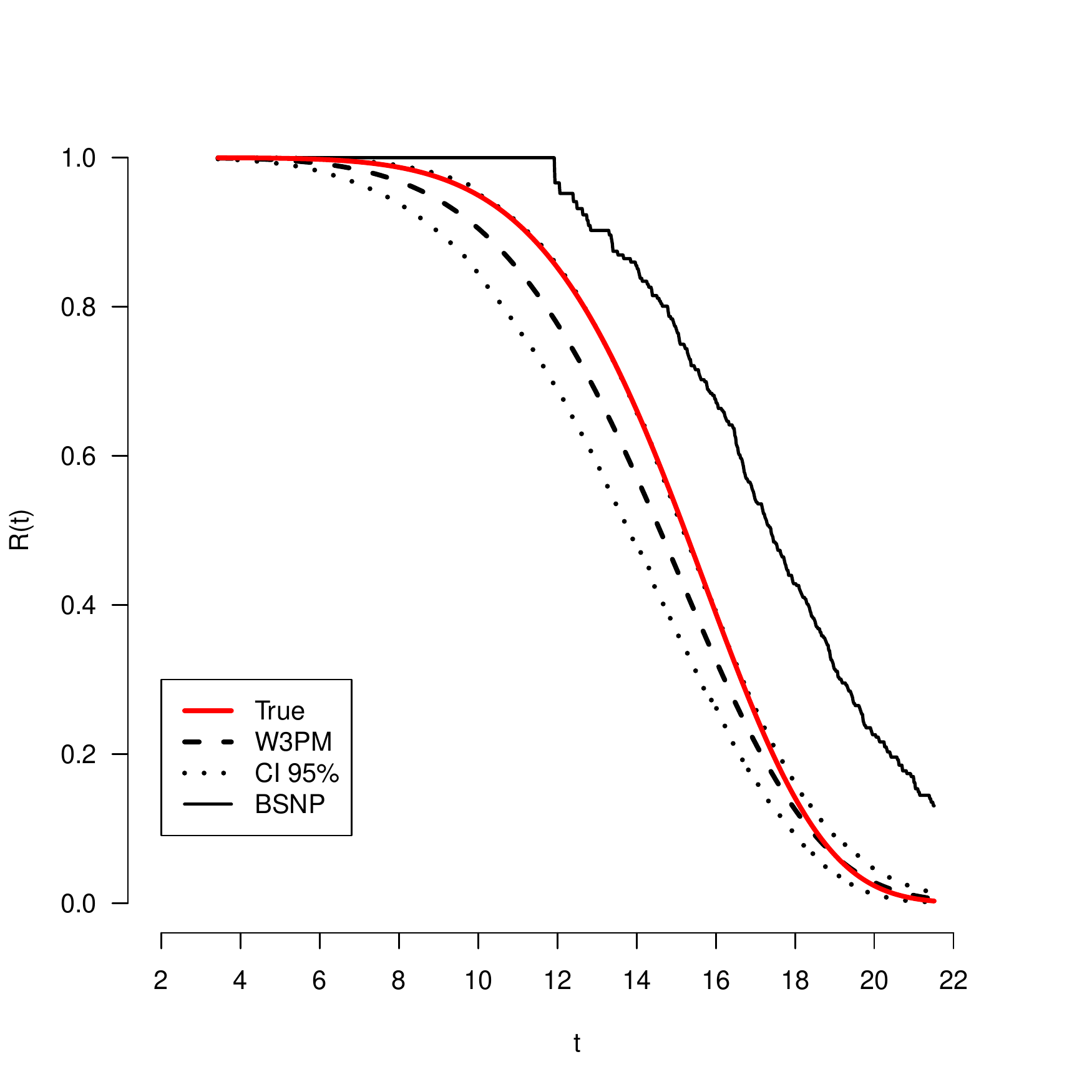}
		\subcaption{Component 1  } \label{ex_simu_2de3_a}
	\end{minipage} 
	\begin{minipage}[b]{0.32\linewidth}
		\includegraphics[width=\linewidth]{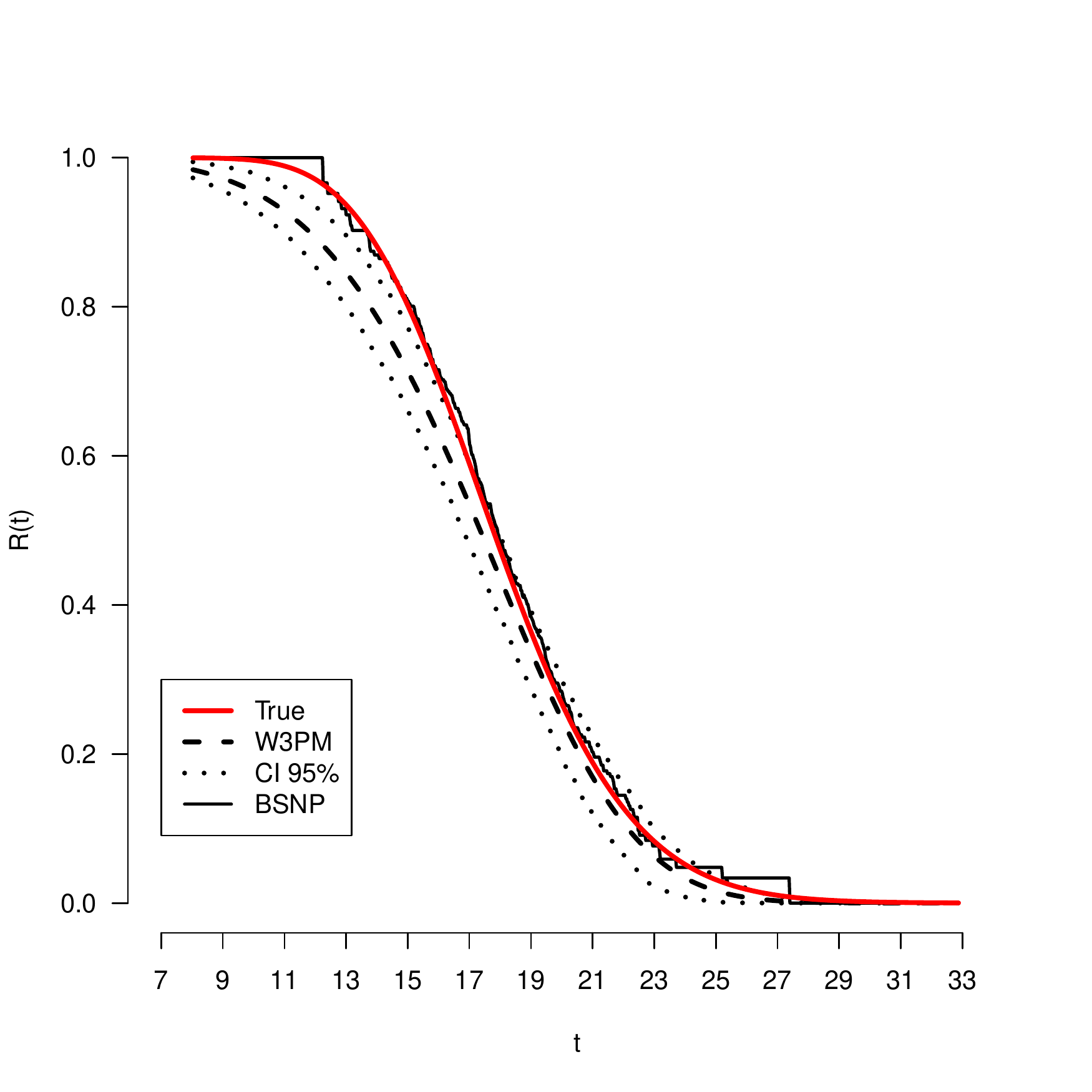}
		\subcaption{Component 2  } \label{ex_simu_2de3_b}
	\end{minipage}
	\begin{minipage}[b]{0.32\linewidth}
		\includegraphics[width=\linewidth]{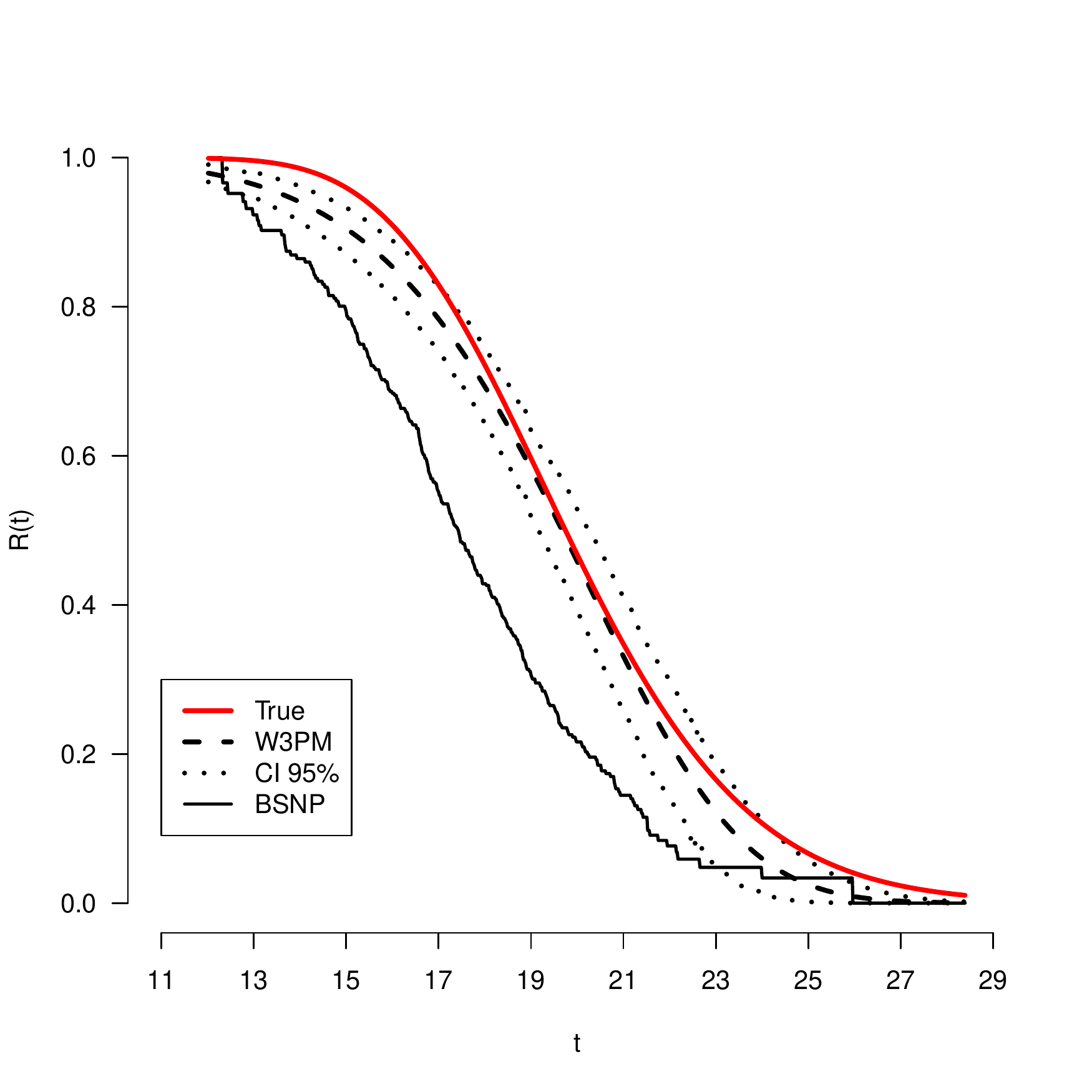}
		\subcaption{Component 3 } \label{ex_simu_2de3_c}
	\end{minipage}	
	\caption{True reliability functions and estimated curves by W3PM and BSNP for the components involved in system structure  1.}
	\label{ex_simu_2de3}
\end{figure}

The MAEs from W3PM and BSNP estimators to the true reliability function are presented in Table \ref{MAE_2de3_simu} and the reliability functions can be visualized in Figure \ref{ex_simu_2de3}, besides the empirical 95\% HPD intervals (CI 95\%) obtained by the proposed model. We can note that W3PM presents reliability curves closer to the true one for components 1 and 3, in which the MAE of posterior mean is much smaller than that obtained by BSNP estimator. BSNP presents lightly less MAE for component 2, however for values of $t$ that the posterior mean is more distant from the true curve, the upper limit of the HPD interval is very close to the true curve, as we can note in Figure \ref{ex_simu_2de3_b}.

\begin{table}[htbp]
	\centering
	\caption{Posterior measures of components reliability functions involved in system structure  1 for some values of $t$.}
	\begin{tabular}{ccccccccc}
		\hline
		\multicolumn{9}{c}{Component 1} \\
		\hline
		t   & Min. & 1st Qu. & Median & Mean & 3rd Qu. & Max. & SD  & HPD 95\%  \\
		3.5 & 0.995 & 0.999 & 1.000 & 0.999 & 1.000 & 1.000 & 0.001 & (0.998;  1.000) \\
		10.0 & 0.788 & 0.889 & 0.908 & 0.905 & 0.927 & 0.966 & 0.029 & (0.842;  0.951) \\
		20.0 & 0.007 & 0.021 & 0.026 & 0.028 & 0.033 & 0.059 & 0.009 & (0.012;  0.046) \\
		\hline
		\multicolumn{9}{c}{Component 2} \\
		\hline
		t   & Min. & 1st Qu. & Median & Mean & 3rd Qu. & Max. & SD  & HPD 95\%  \\
		8.5 & 0.938 & 0.975 & 0.980 & 0.979 & 0.984 & 0.994 & 0.007 & (0.965;  0.992) \\
		15.0 & 0.602 & 0.692 & 0.712 & 0.712 & 0.732 & 0.793 & 0.030 & (0.660;   0.772) \\
		26.0 & 0.001 & 0.003 & 0.006 & 0.007 & 0.009 & 0.056 & 0.006 & (0.001;  0.020) \\
		\hline
		\multicolumn{9}{c}{Component 3} \\
		\hline
		t   & Min. & 1st Qu. & Median & Mean & 3rd Qu. & Max. & SD  & HPD 95\%  \\
		12.5 & 0.936 & 0.968 & 0.974 & 0.973 & 0.978 & 0.989 & 0.008 & (0.958;  0.986) \\
		20.0 & 0.347 & 0.437 & 0.460 & 0.458 & 0.480 & 0.555 & 0.034 & (0.392;  0.528) \\
		25.0 & 0.001 & 0.012 & 0.021 & 0.024 & 0.033 & 0.119 & 0.016 & (0.002;  0.056) \\
		\hline
	\end{tabular}%
	\label{medidas_posterior_2de3}%
\end{table}%

\begin{table}[h!]
	\centering
	\caption{MAE of W3PM and BSNP estimators when compared to true reliability functions for  system structure  1.}
	\begin{tabular}{cccc}
		\hline
		  & \multicolumn{3}{c}{Component } \\
		& 1 &  2 &  3  \\
		\hline
		W3PM  & 0.0350 & 0.0282 & 0.0259  \\
		BSNP  & 0.1356 & 0.0176 & 0.1332  \\
		\hline
	\end{tabular}%
	\label{MAE_2de3_simu}%
\end{table}%

\newpage

\subsection{System 2}

\begin{figure}[h!]\centering
	\begin{minipage}[b]{0.32\linewidth}
		\includegraphics[width=\linewidth]{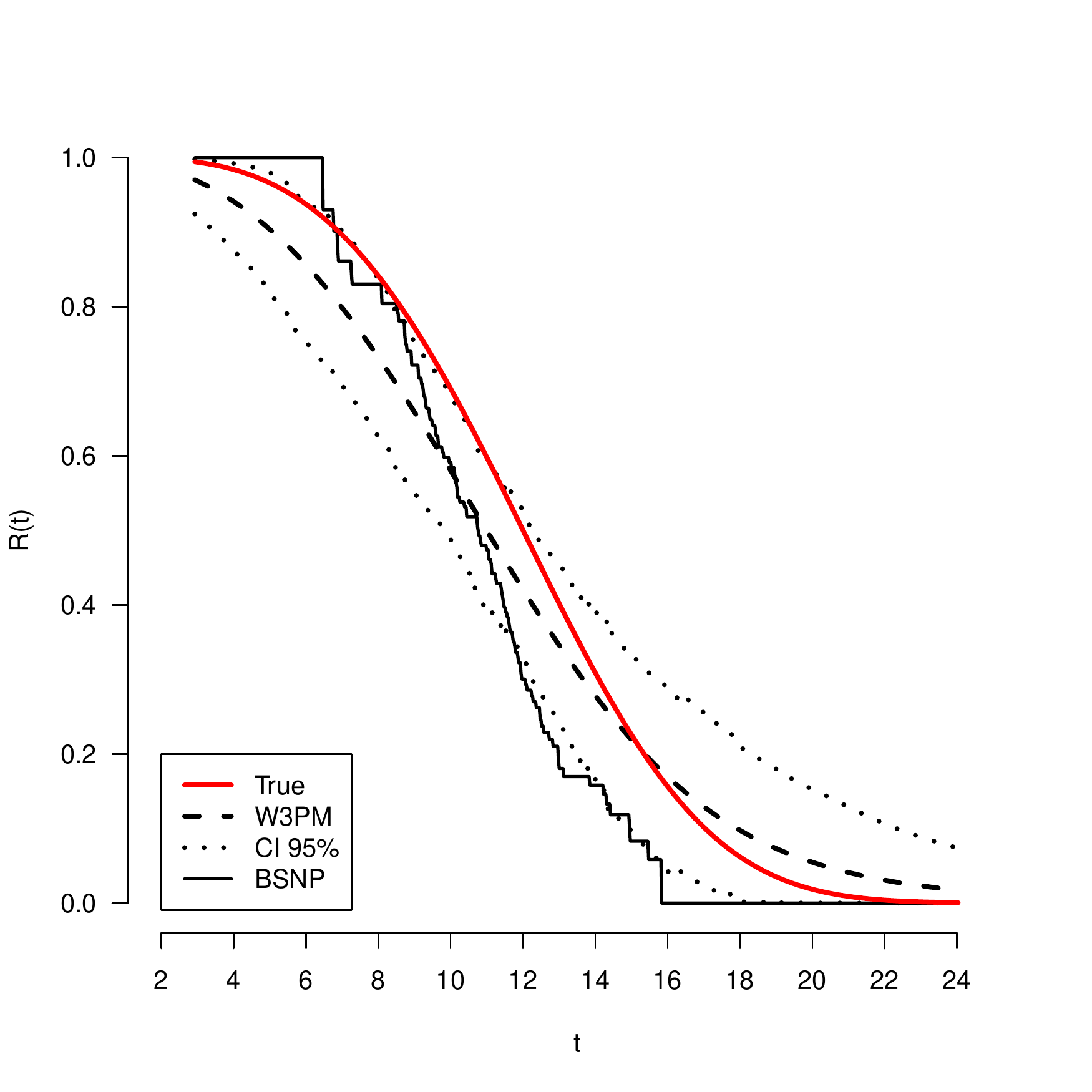}
		\subcaption{Component 1 } \label{ex_simu_chin_a}
	\end{minipage} 
	\begin{minipage}[b]{0.32\linewidth}
		\includegraphics[width=\linewidth]{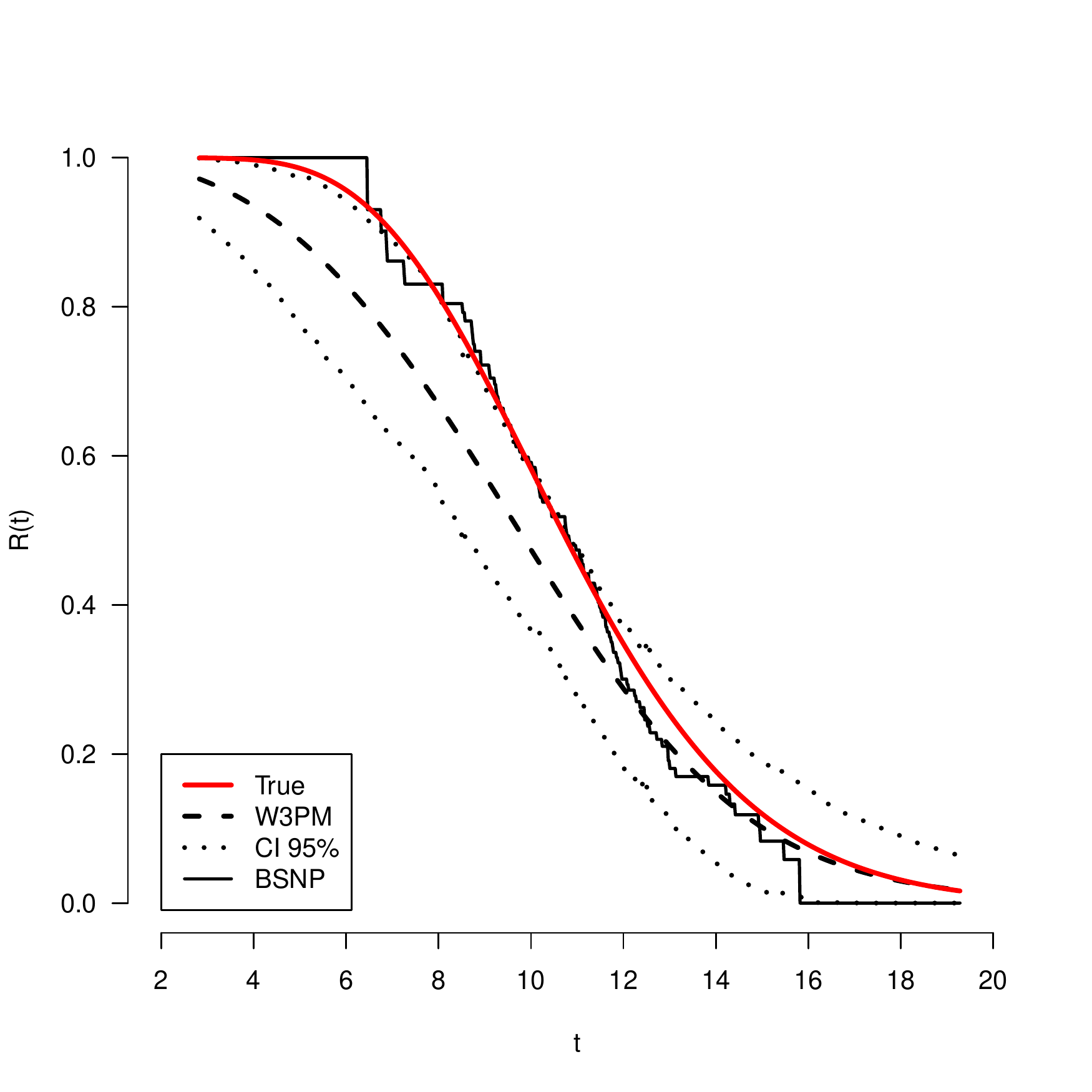}	
		\subcaption{Component 2  } \label{ex_simu_chin_b}
	\end{minipage}
	\begin{minipage}[b]{0.32\linewidth}
		\includegraphics[width=\linewidth]{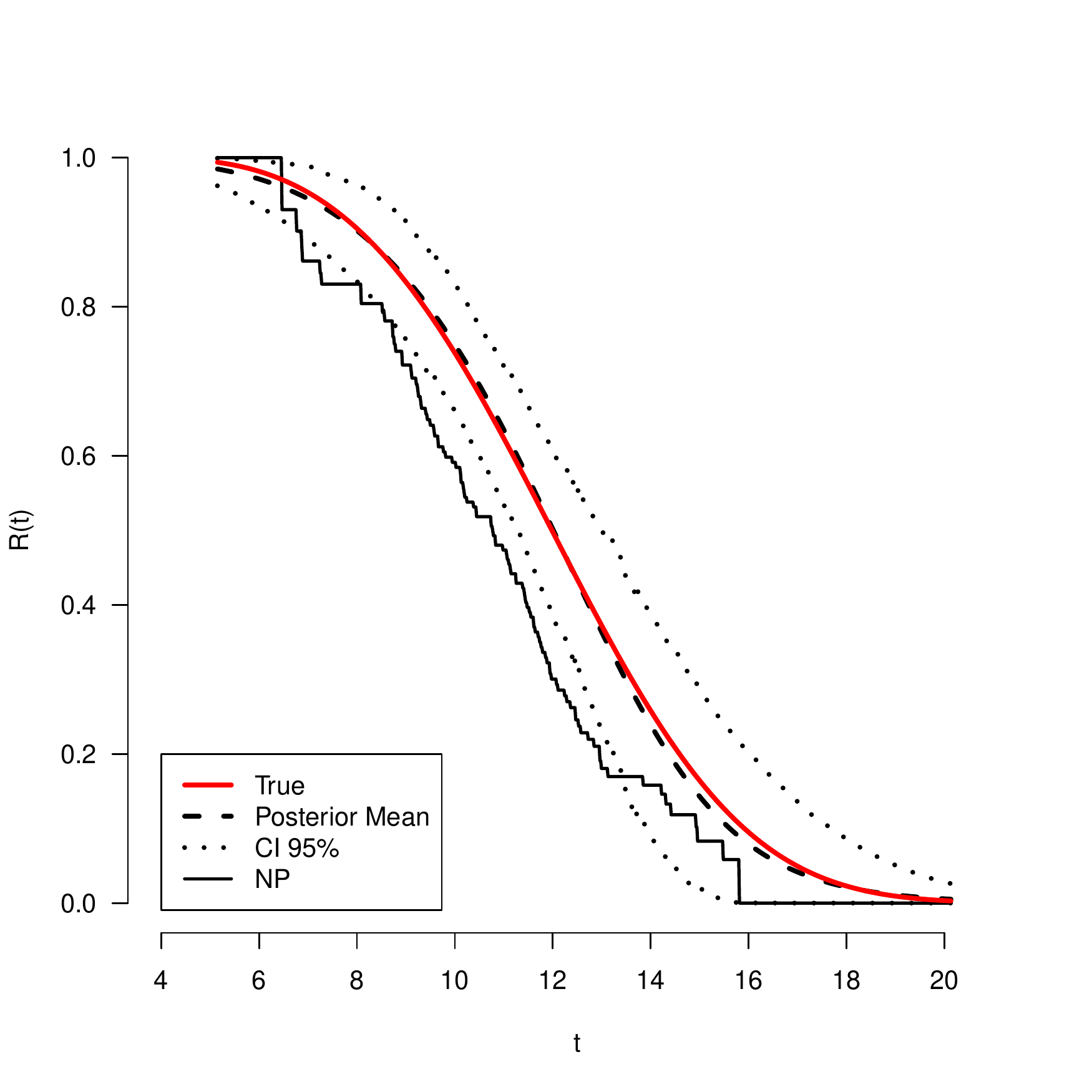}
		\subcaption{Component 3 } \label{ex_simu_chin_c}
	\end{minipage}	
	\begin{minipage}[b]{0.32\linewidth}
		\includegraphics[width=\linewidth]{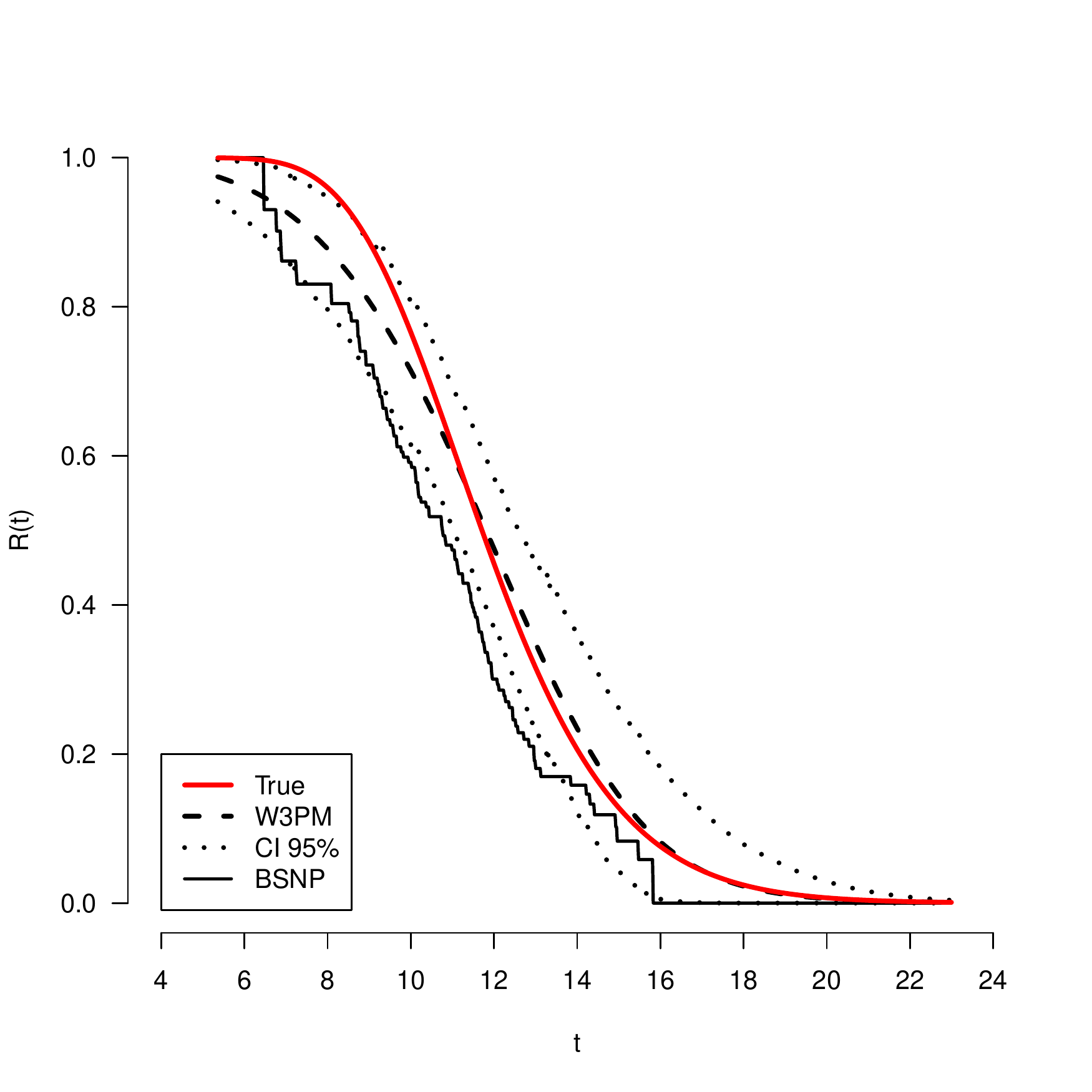}
		\subcaption{Component 4 } \label{ex_simu_chin_d}
	\end{minipage}	
	\begin{minipage}[b]{0.32\linewidth}
		\includegraphics[width=\linewidth]{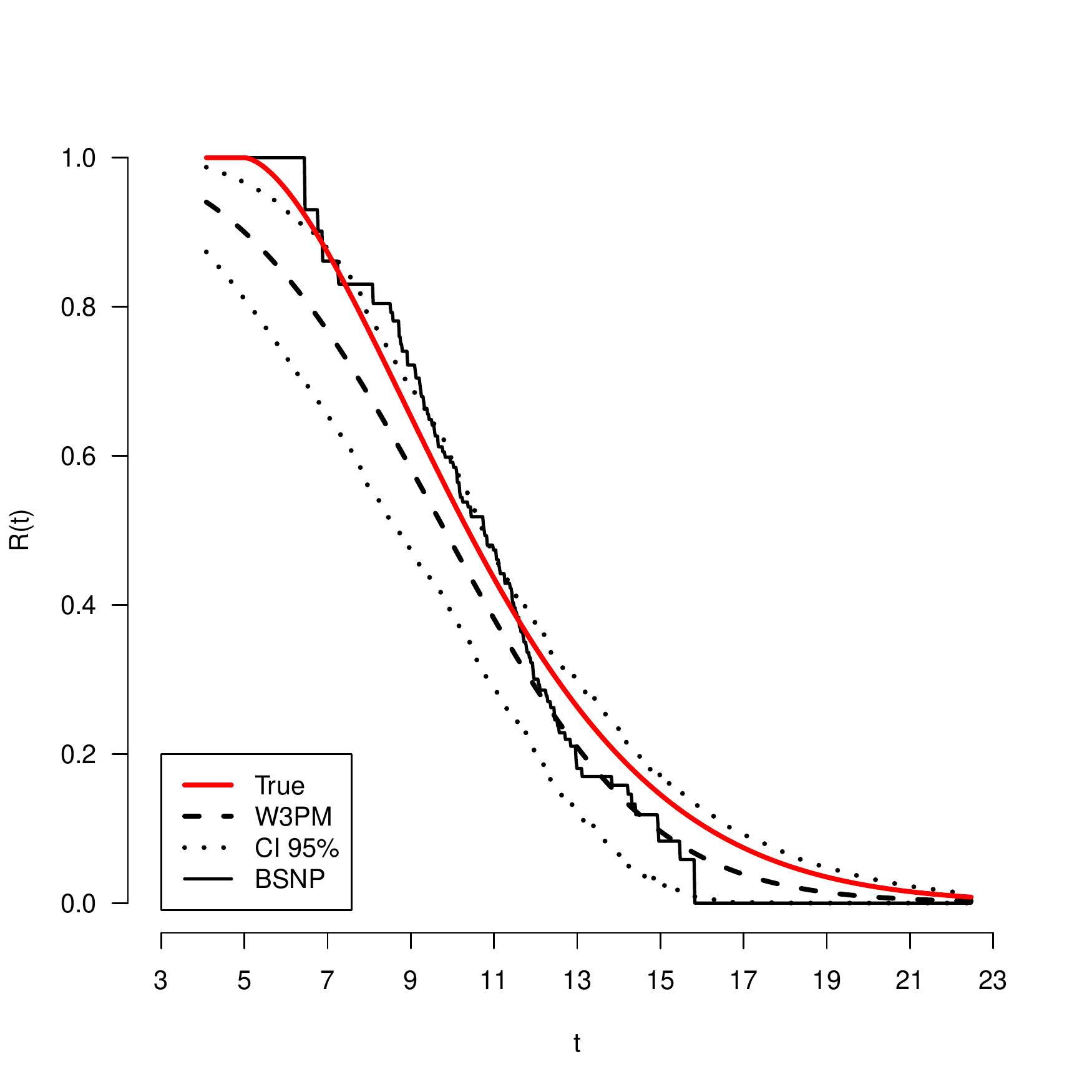}
		\subcaption{Component 5 } \label{ex_simu_chin_e}
	\end{minipage}	
	\caption{True reliability functions and estimated curves by W3PM and BSNP for the components involved in system structure 2.}
	\label{ex_simu_chin}
\end{figure}

In Table \ref{MAE_chin_simu} we present the MAE values obtained by W3PM and BSNP in relation to the true reliability function for each of the five components. The true reliability functions, posterior means, 95\% HPD intervals (CI 95\%) and the BSNP estimates can be visualized in Figure \ref{ex_simu_chin}. The proposed model presents lower MAE values for components 1, 3 and 4, in which the proposed model presents an almost perfect reliability function estimation for the component 3. Even for components 2 and 5, in which the BSNP estimator presented slightly lower MAE, the 95\% HPD interval contains the true curves for almost all values of $t$, as we can note in Figure \ref{ex_simu_chin_b} and \ref{ex_simu_chin_e}.

\begin{table}[htbp]
	\centering
	\caption{MAE of W3PM and BSNP estimators when compared to true reliability functions for system structure 2.}
	\begin{tabular}{cccccc}
		\hline
		   & \multicolumn{5}{c}{Component } \\
		&  1 & 2 &  3 &  4 &  5  \\
		\hline
		W3PM & 0.0426 & 0.0581 & 0.0180 & 0.0381 & 0.0390  \\
		BSNP  & 0.0583 & 0.0343 & 0.0750 & 0.0538 & 0.0321  \\
		\hline
	\end{tabular}%
	\label{MAE_chin_simu}%
\end{table}%

\newpage

\subsection{System 3} \label{ex_simu_bri_texto}

\begin{figure}[h!]\centering
	\begin{minipage}[b]{0.32\linewidth}
		\includegraphics[width=\linewidth]{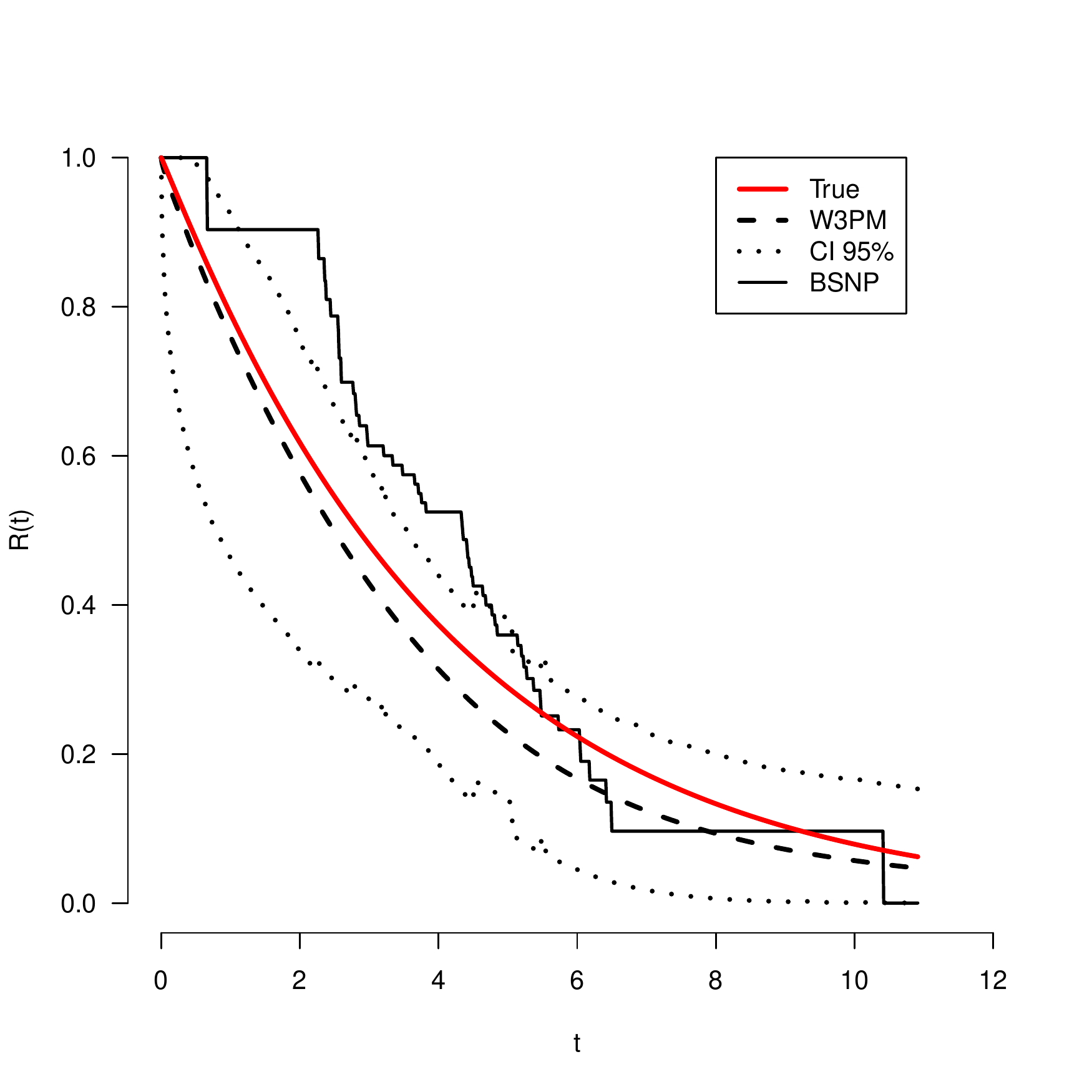}
		\subcaption{Component 1  }
	\end{minipage} 
	\begin{minipage}[b]{0.32\linewidth}
		\includegraphics[width=\linewidth]{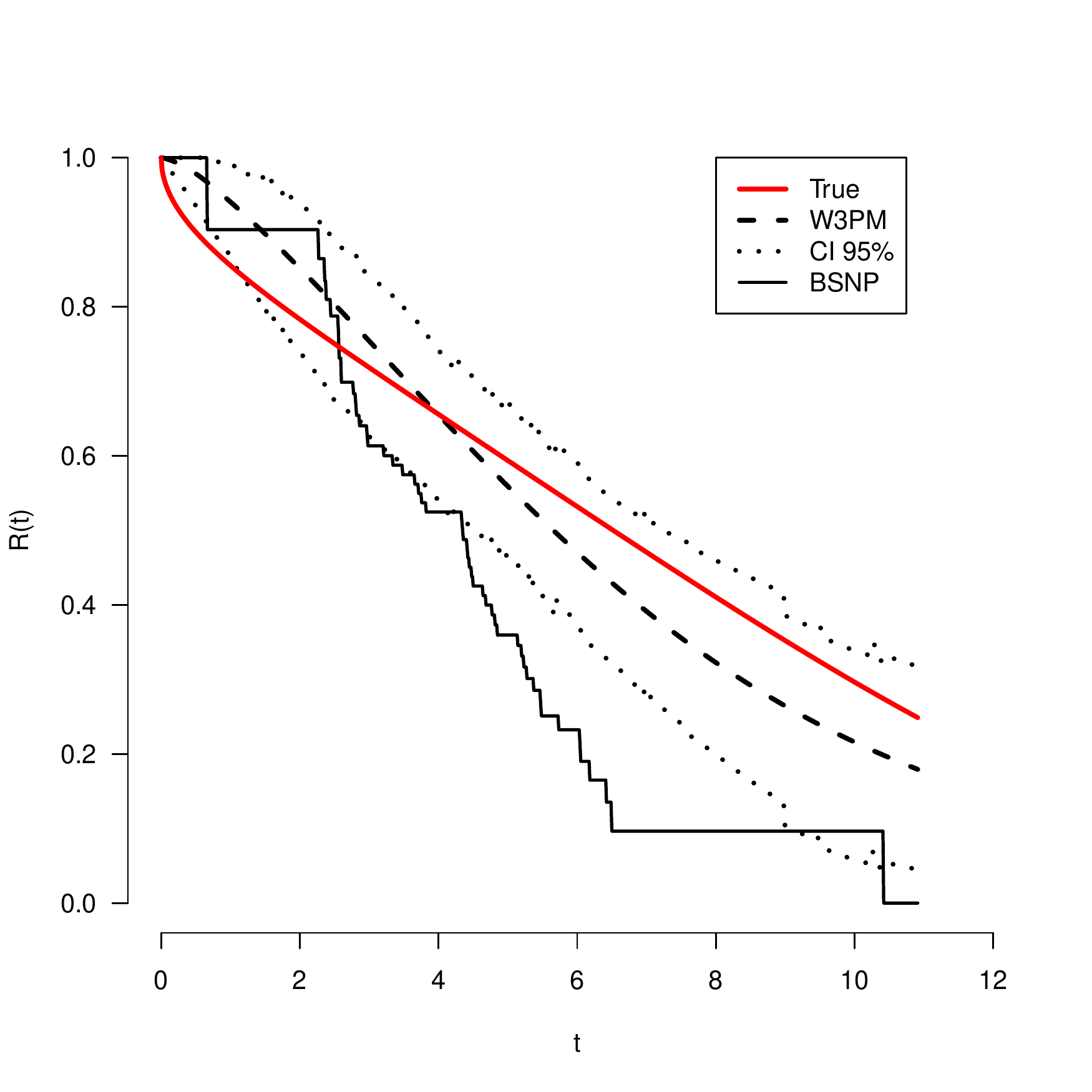}
		\subcaption{Component 2  }
	\end{minipage}
	\begin{minipage}[b]{0.32\linewidth}
		\includegraphics[width=\linewidth]{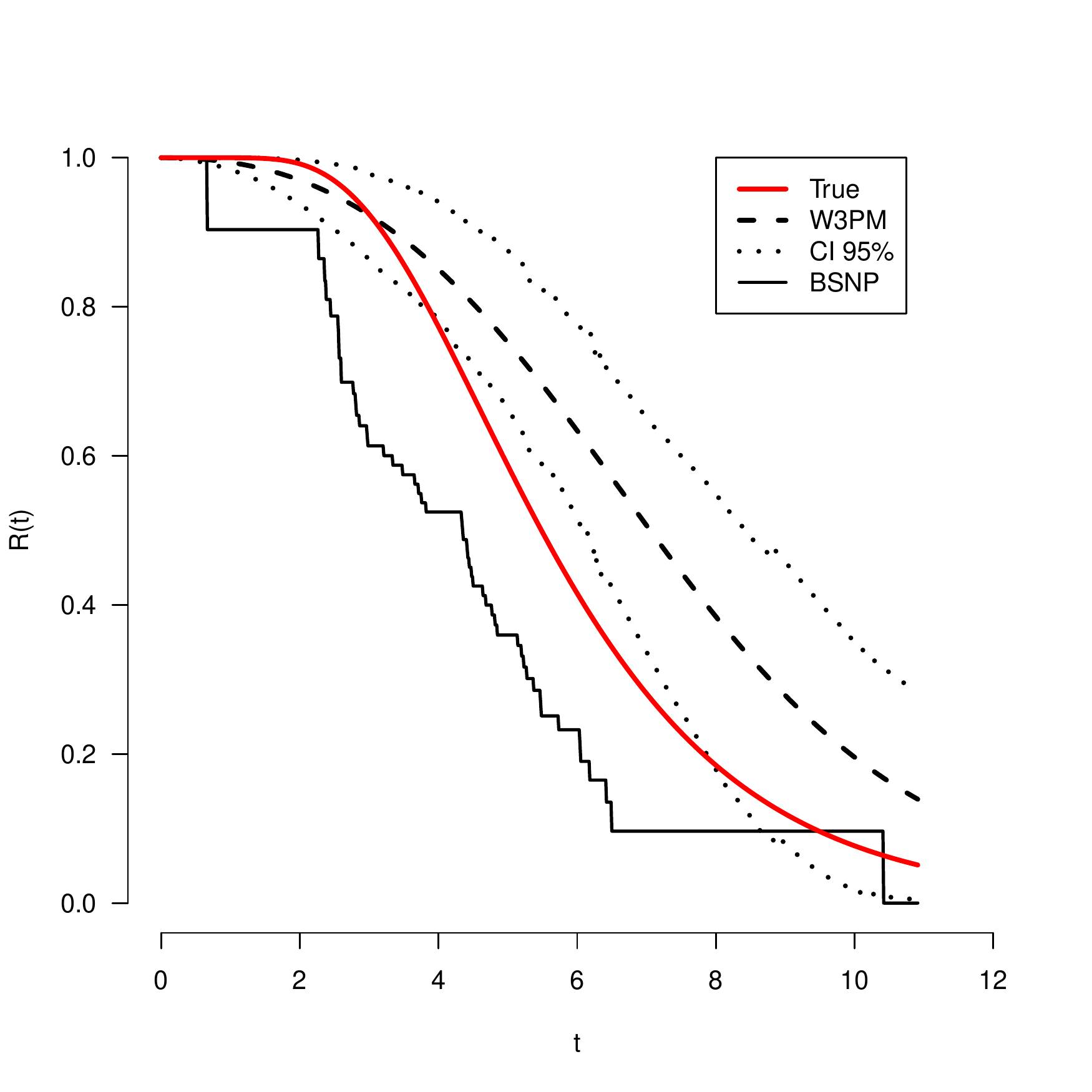}
		\subcaption{Component 3 }
	\end{minipage}	
	\begin{minipage}[b]{0.32\linewidth}
		\includegraphics[width=\linewidth]{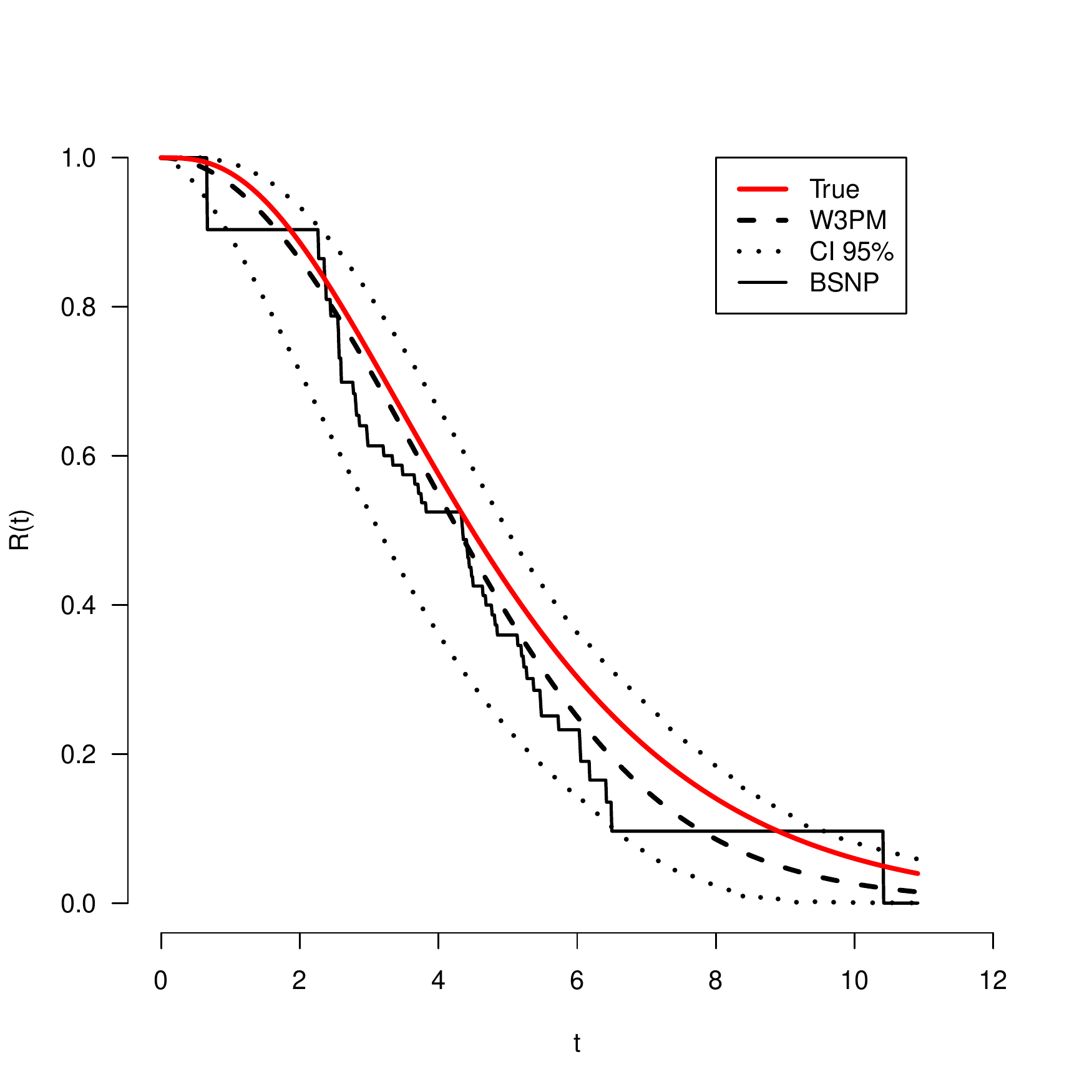}
		\subcaption{Component 4 }
	\end{minipage}	
	\begin{minipage}[b]{0.32\linewidth}
		\includegraphics[width=\linewidth]{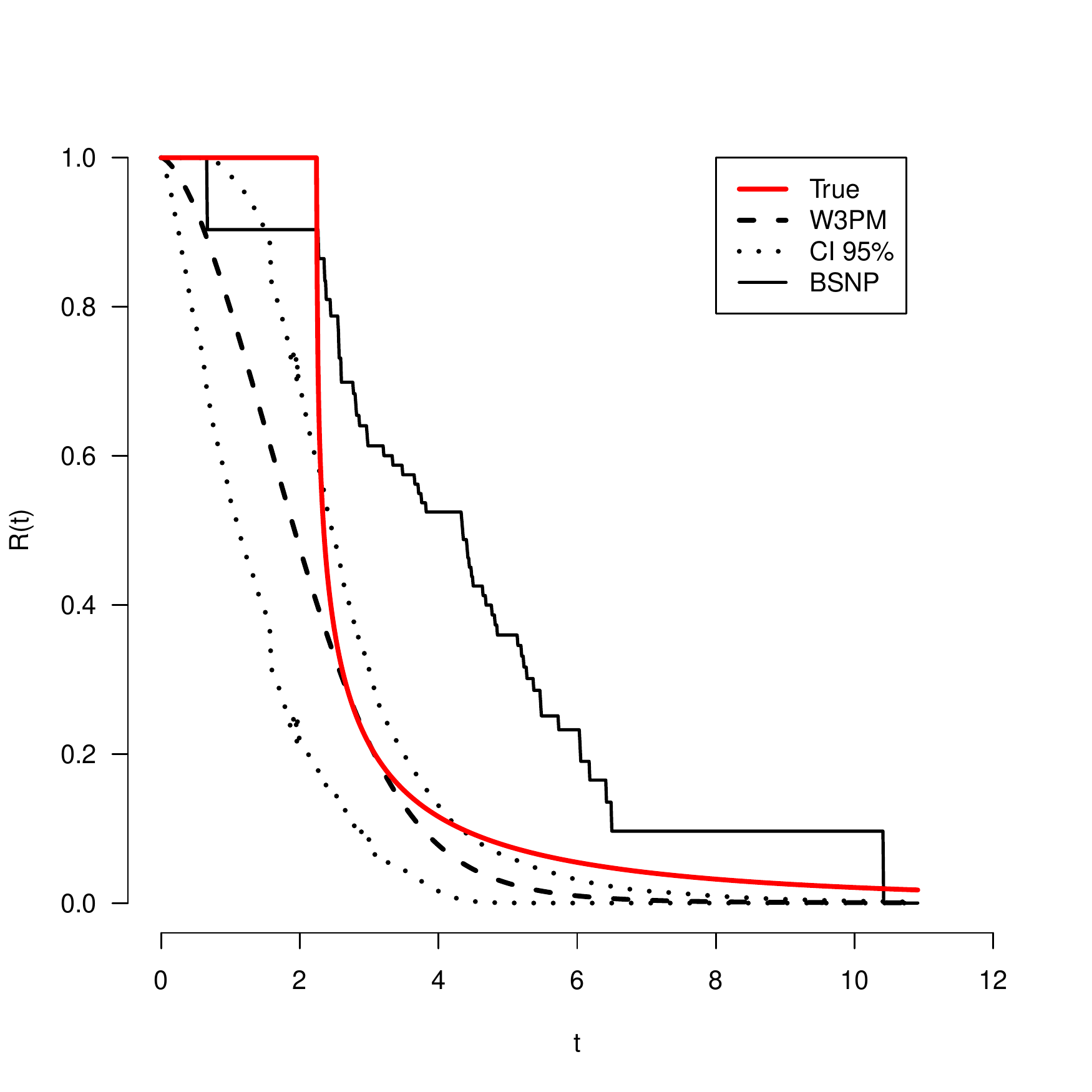}
		\subcaption{Component 5 }
	\end{minipage}	
	\caption{True reliability functions and estimated by W3PM and BSNP for the components involved system structure 3.}
	\label{ex_simu_bri}
\end{figure}

In Table \ref{MAE_bri_simu} we present the MAE values obtained by W3PM and BSNP in relation to the true reliability functions and in Figure \ref{ex_simu_bri} there are the true curves, posterior means, 95\% HPD intervals (CI 95\%) and the BSNP estimates. The proposed model presents lower MAE values for all five components involved in the simulated bridge system.

\begin{table}[h!]
	\centering
	\caption{MAE of W3PM and BSNP estimators when compared to true reliability functions for system structure 3.}
	\begin{tabular}{cccccc}
		\hline
		     & \multicolumn{5}{c}{Component }  \\
		&   1 &  2 &  3 &  4 &   5  \\
		\hline
		W3PM   & 0.0384 & 0.0758 & 0.0976 & 0.0365  & 0.0565  \\
		BSNP  & 0.0671 & 0.0936 & 0.1193 & 0.0505 & 0.1777  \\
		\hline
	\end{tabular}%
	\label{MAE_bri_simu}%
\end{table}%

\section{Model Evaluation with Simulation Studies} \label{simulation_study}

To evaluate the performance of the proposed model, this section presents simulation studies  with different sample sizes, system structure and proportion of masked data. The proposed model under symmetric assumption and considering $\lambda_{2j}=0$ is compared to the BSNP estimator, which has been seen as the best choice for masked data in complex systems. 

Two types of system structures are used: $2$-out-of-$3$ (Figure \ref{2de3}) and bridge system (Figure \ref{bridge}). Four sample sizes are considered ($n=50,~100,~300, ~1000$) and three proportions of masked data: $p=0.2, ~ 0.4$ and $0.7$. For each scenario (combination of sample size and proportion of masked data), $1000$ samples were generated from systems presented in \ref{ex_simu_2de3_texto} and in \ref{ex_simu_bri_texto}. Mean absolute error (MAE) from the estimators to the true distribution is considered as the comparison measure.

\begin{figure}[h!]\centering
	\begin{minipage}[b]{0.32\linewidth}
		\includegraphics[width=\linewidth]{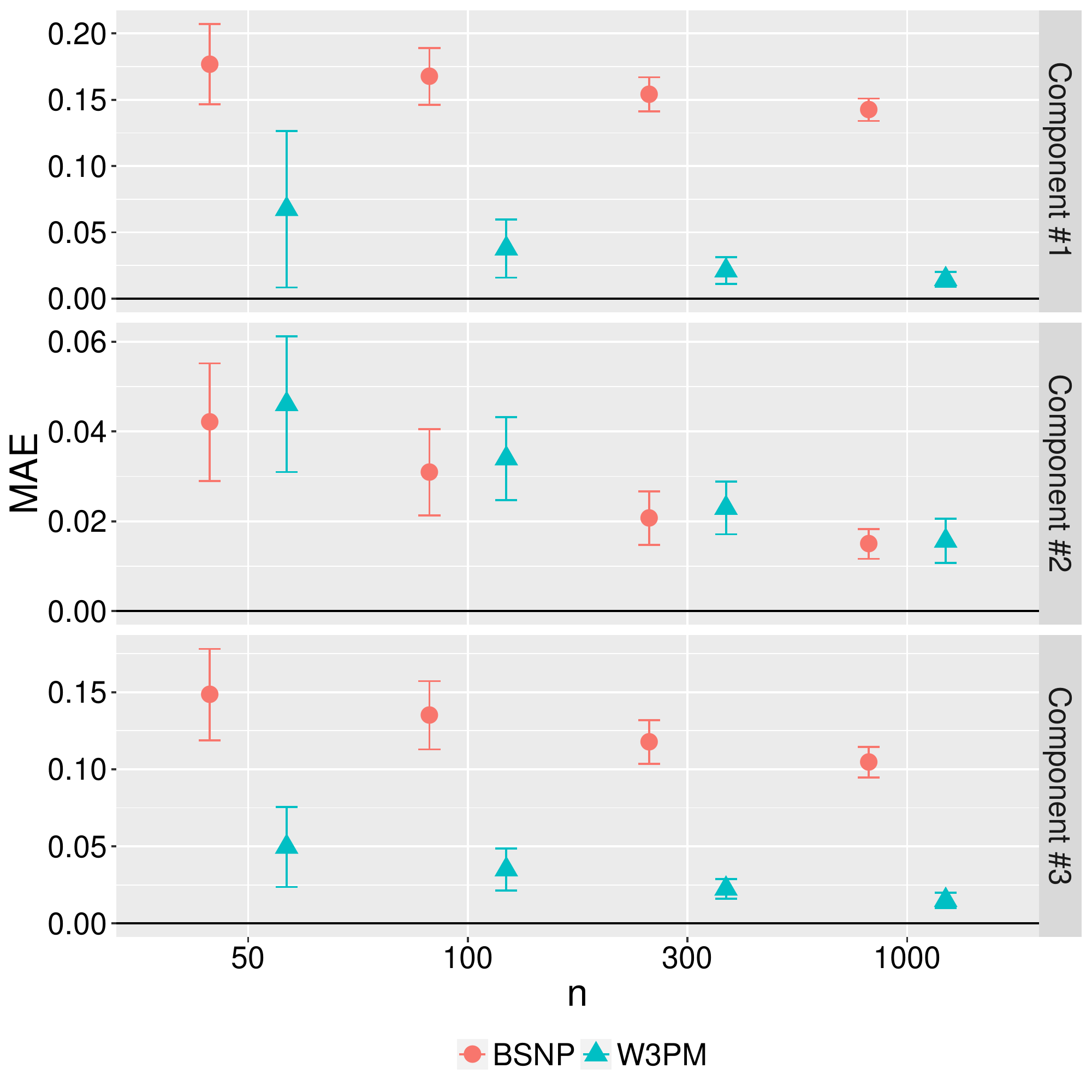}
		\subcaption{$p=0.2$  }
	\end{minipage} 
	\begin{minipage}[b]{0.32\linewidth}
		\includegraphics[width=\linewidth]{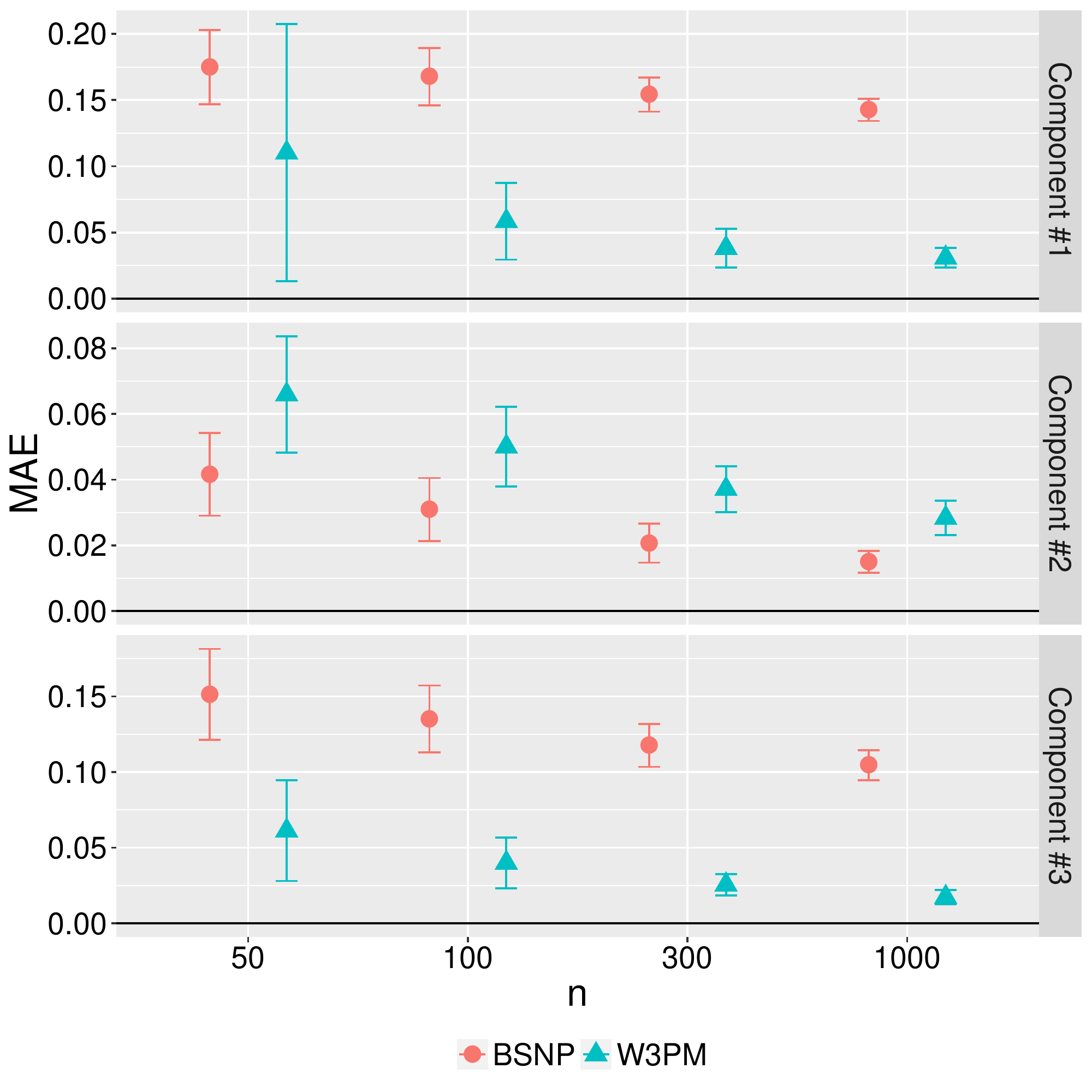}
		\subcaption{$p=0.4$  }
	\end{minipage}
	\begin{minipage}[b]{0.32\linewidth}
		\includegraphics[width=\linewidth]{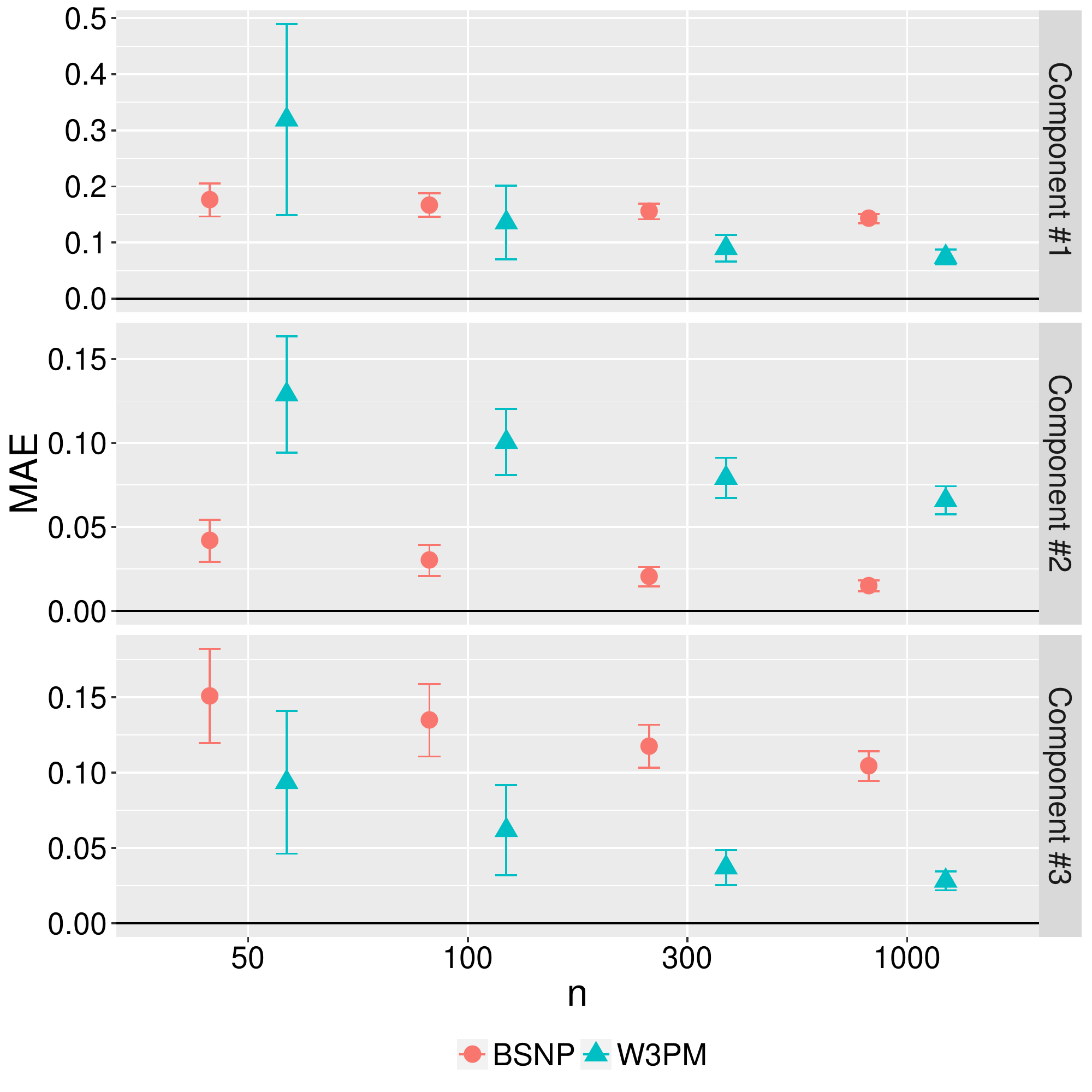}
		\subcaption{$p=0.7$ }
	\end{minipage}	
	\caption{Mean (symbol) and standard deviation (bars) of MAE obtined by W3PM and BSNP for $2$-out-of-$3$ structure.}
	\label{simu_2de3}
\end{figure}

\begin{figure}[h!]\centering
	\begin{minipage}[b]{0.32\linewidth}
		\includegraphics[width=\linewidth]{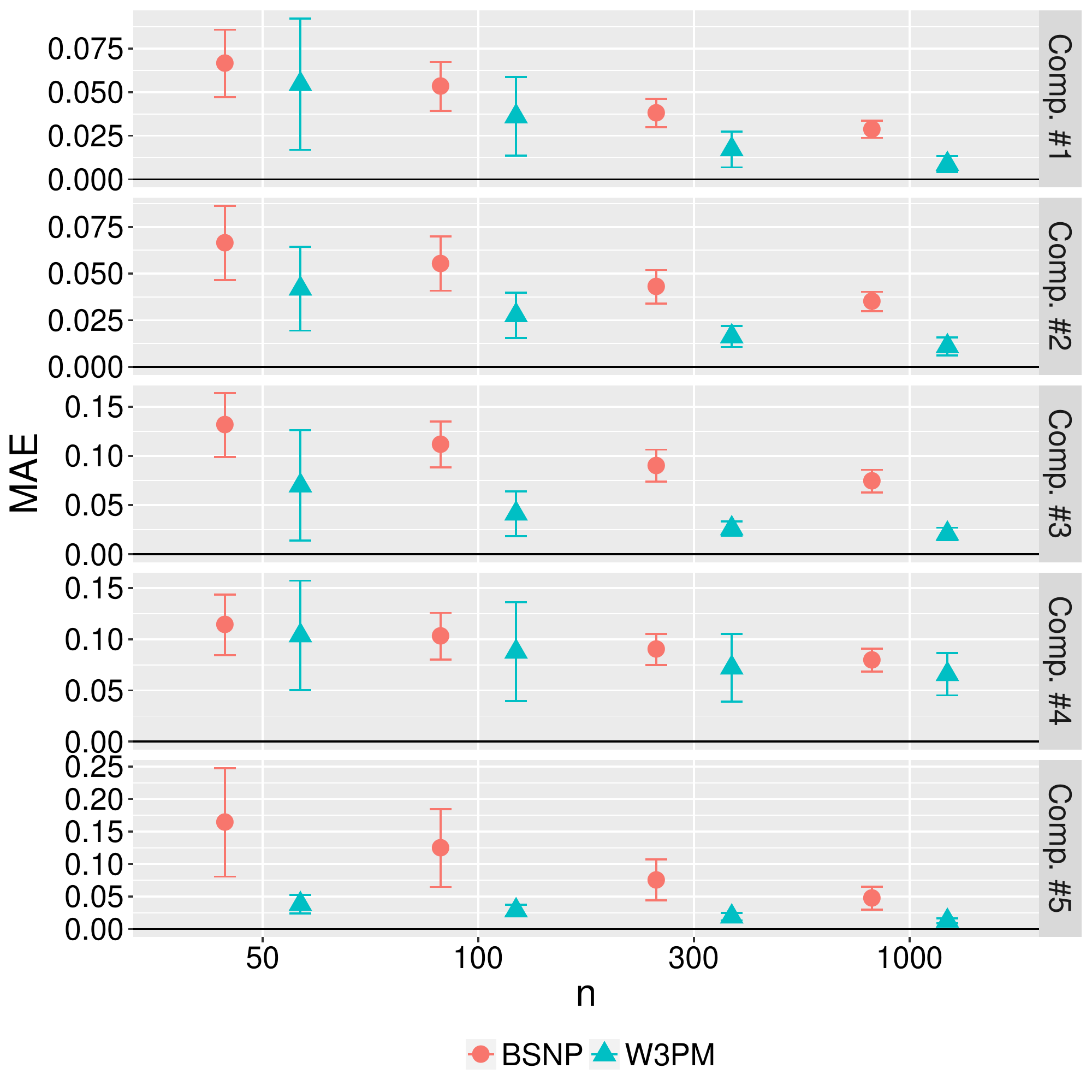}
		\subcaption{$p=0.2$  }
	\end{minipage} 
	\begin{minipage}[b]{0.32\linewidth}
		\includegraphics[width=\linewidth]{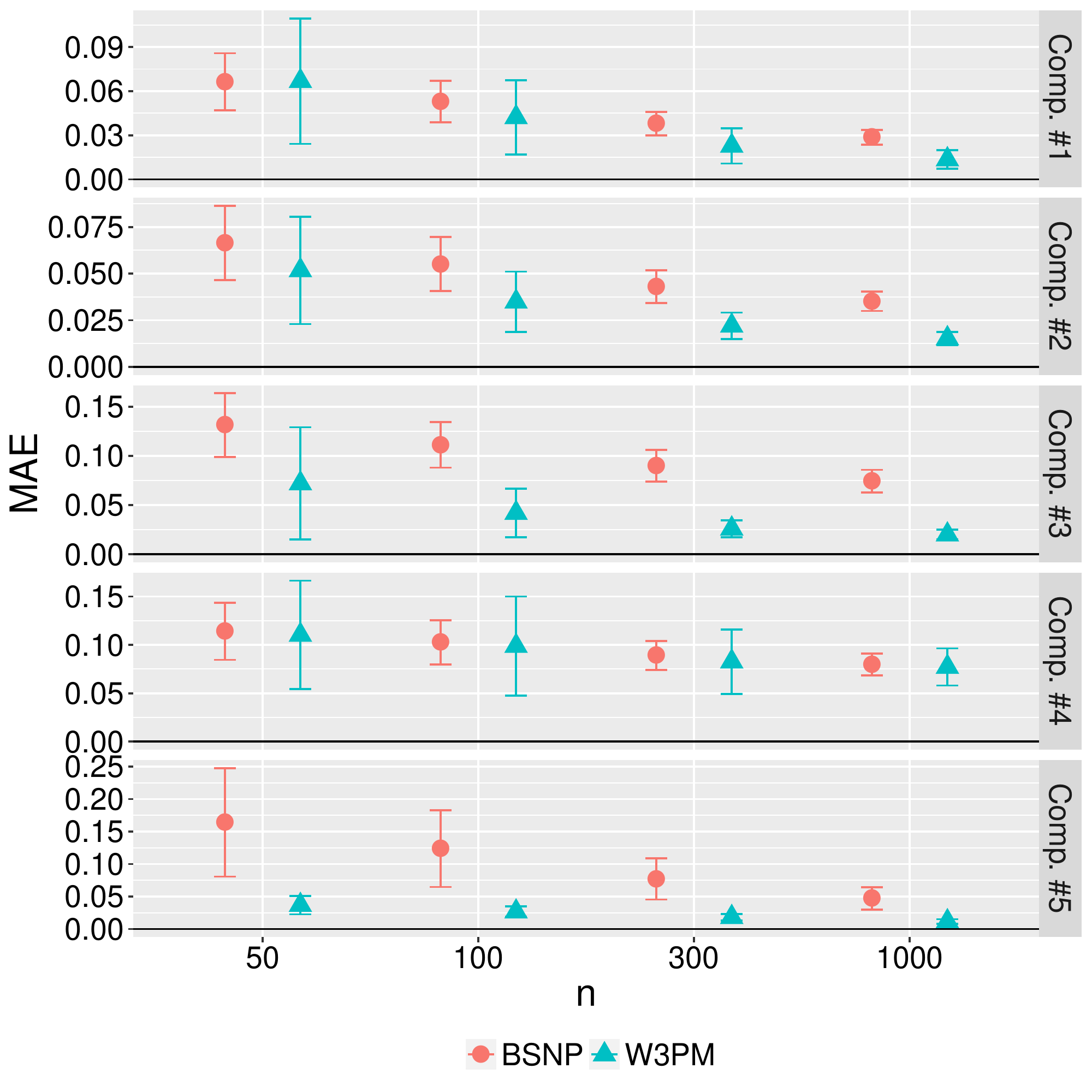}
		\subcaption{$p=0.4$  }
	\end{minipage}
	\begin{minipage}[b]{0.32\linewidth}
		\includegraphics[width=\linewidth]{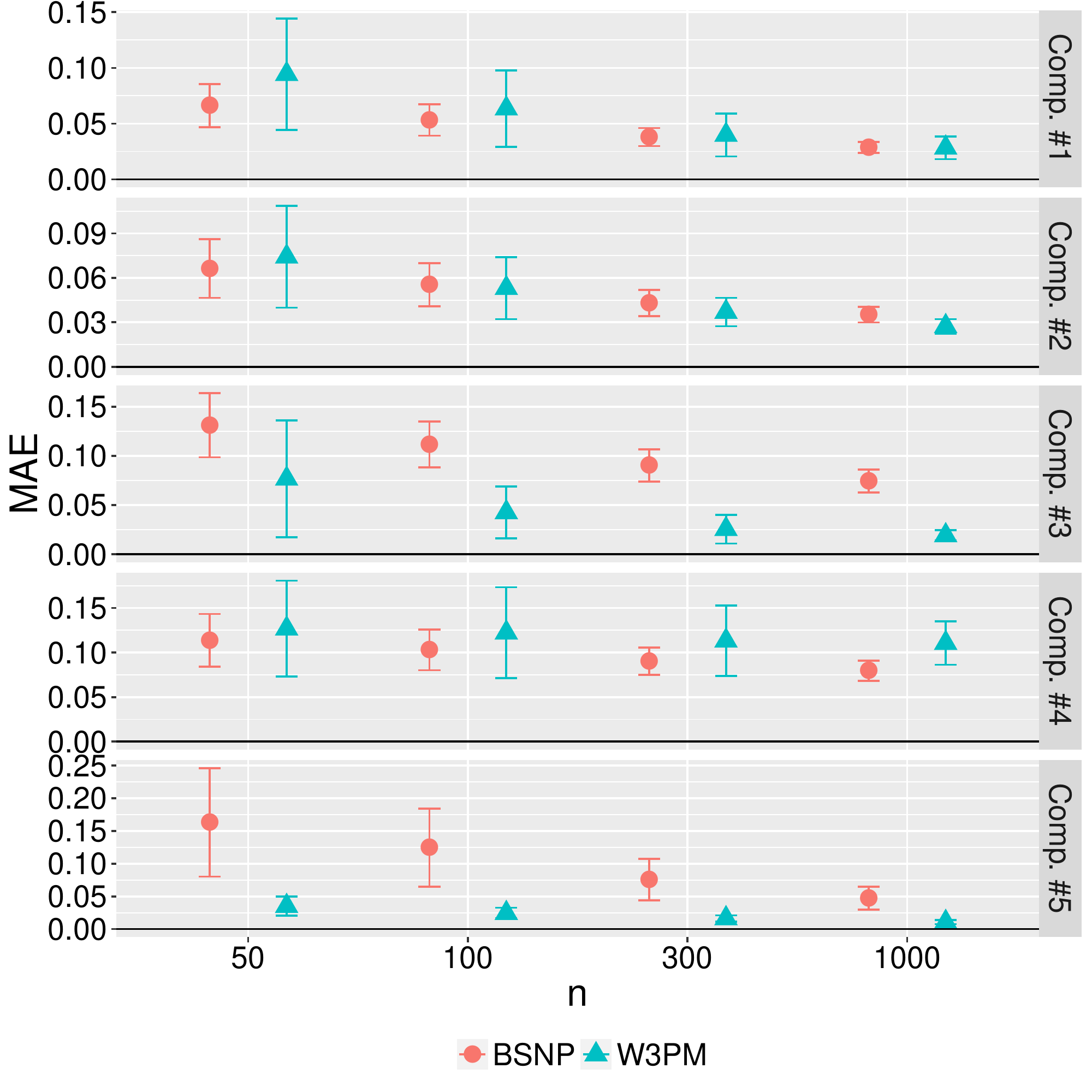}
		\subcaption{$p=0.7$ }
	\end{minipage}	
	\caption{Mean (symbol) and standard deviation (bars) of MAE obtined by W3PM and BSNP for bridge structure.}
	\label{simu_bridge}
\end{figure}

The mean and stardard deviation of $1000$ MAE values obtained by W3PM and BSNP estimator are presented in Figures \ref{simu_2de3} and \ref{simu_bridge} for systems $2$-out-of-$3$ and bridge, respectively. In general, the W3PM presents lower MAE values mean. The exception is for component 2 of the $2$-out-of-$3$ system in which the BSNP estimator presents better performance. However, the difference between the two methods decreases as the sample size increases, mainly because the performance of the proposed estimator improves as $n$ increases. 

%


\section{Aplication} \label{aplication}
In this section, a real dataset is considered in order to present the applicability of the proposed model for reliability estimation of components involved in coherent systems. The dataset is available in \cite{Flehinger2002} which consists of $172$ observed failure times of computer hard-drives monitored over a period of $4$ years. There were three possible causes of failure: eletronic hard (component $j=1$), head flyability (component $j=2$) and head/disc magnetics (component $j=3$). However, for some of them ($38\%$) the cause of hard-drive fail was not identified. For these masked data systems, $\Upsilon=\{1,3\}$ or $\Upsilon=\{1,2,3\}$, that is, there is no possible masked set $\Upsilon=\{1,2\}$ or 
$\Upsilon=\{2,3\}$. Note that in our proposal approach the configuration of set $\Upsilon$ is not a big deal, once the important information for estimation of $j$-th component reliability is if $j$ belongs to $\Upsilon$ or not. More details about the detection of failure causes can be found in \cite{Flehinger} and in \cite{CraiuReiser}.

As we can see from Table \ref{casa_falha_drive}, the component 1 is observed to cause the failure of $20.35\%$ of systems, $11.05\%$ of systems had observed failures because of component 2 and component 3 was known responsible for $30.23\%$ of system failure. Besides, or component 1 or component 3 caused the failure of $18.60\%$ of systems and the remaining $19.77\%$ of systems broken down because of any of three components.

Since the components in $\Upsilon$ are right-censored or responsible for system failure, $\lambda_{3j}=0$, for $j=1,2,3$, and no hard-drive is subject of left-censored failure time.

We fitted the proposed model under and not under the symmetric assumption and the BSNP estimates are obtained. 
For proposed model, we generated $35000$ values of each parameter, disregarding the first $5000$ iterations to eliminate the effect of the initial values and spacing of size $30$ to avoid correlation problems, obtaining a sample of size $n_p=1000$. The chains convergence was monitored and good convergence results were obtained.

The proposed model parameters estimates are presented in Table \ref{post_hard_drive}. The two versions of proposed model presents close estimates for $\bm{\theta}=(\beta,\eta,\mu)$ for all components. It is worth mentioning that the posterior mean of $\mu$ is close to zero for three failure causes, which indicates that the beginning of computer lifetimes coincides with the beginning of experiment, which makes sense as it is a controlled experiment that hard-drives were not tested previously. 

\begin{figure}[h!]\centering
	\begin{minipage}[b]{0.3\linewidth}
		\includegraphics[width=\linewidth]{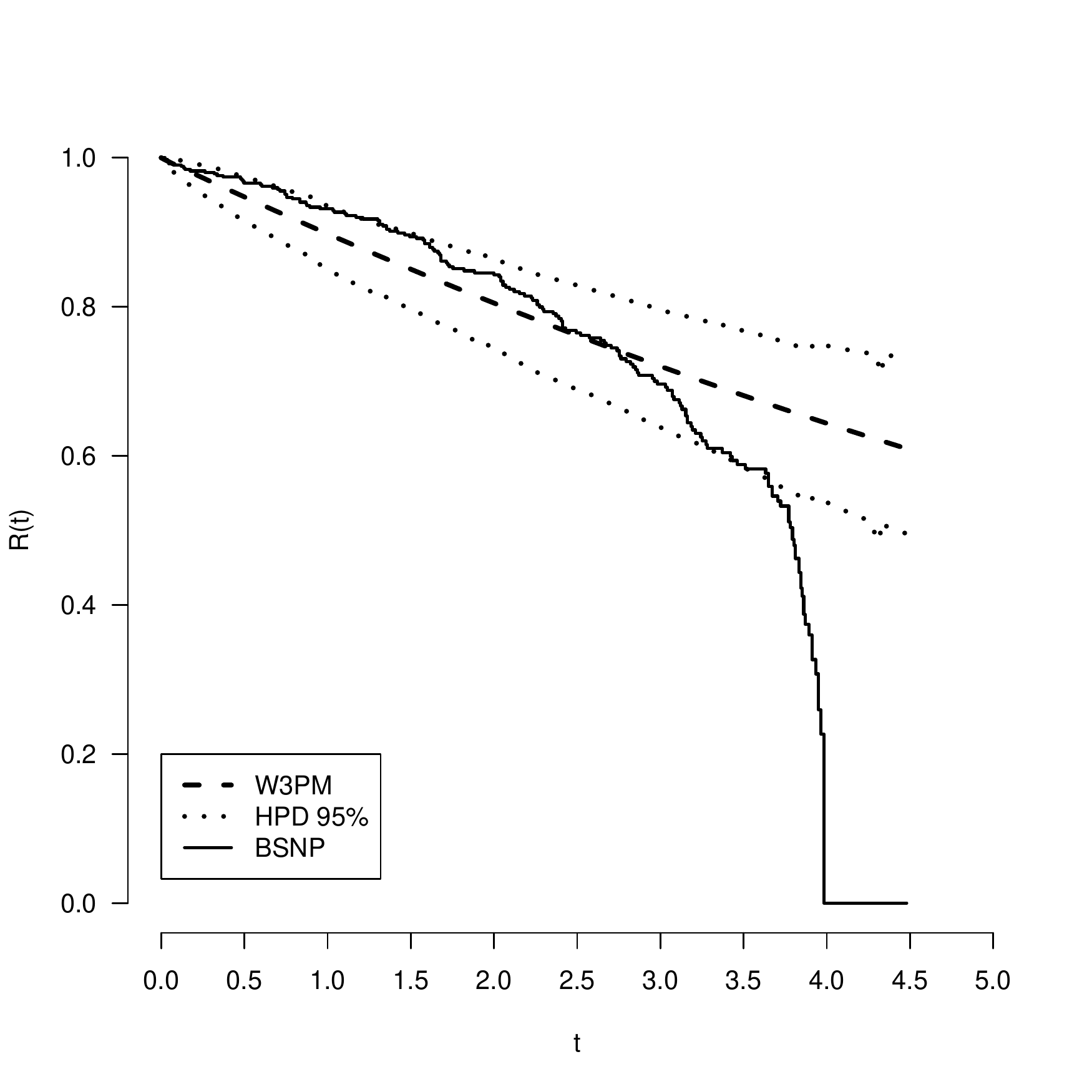}
		\subcaption{Eletronic hard ($j=1$)}
	\end{minipage} 
	\begin{minipage}[b]{0.3\linewidth}
		\includegraphics[width=\linewidth]{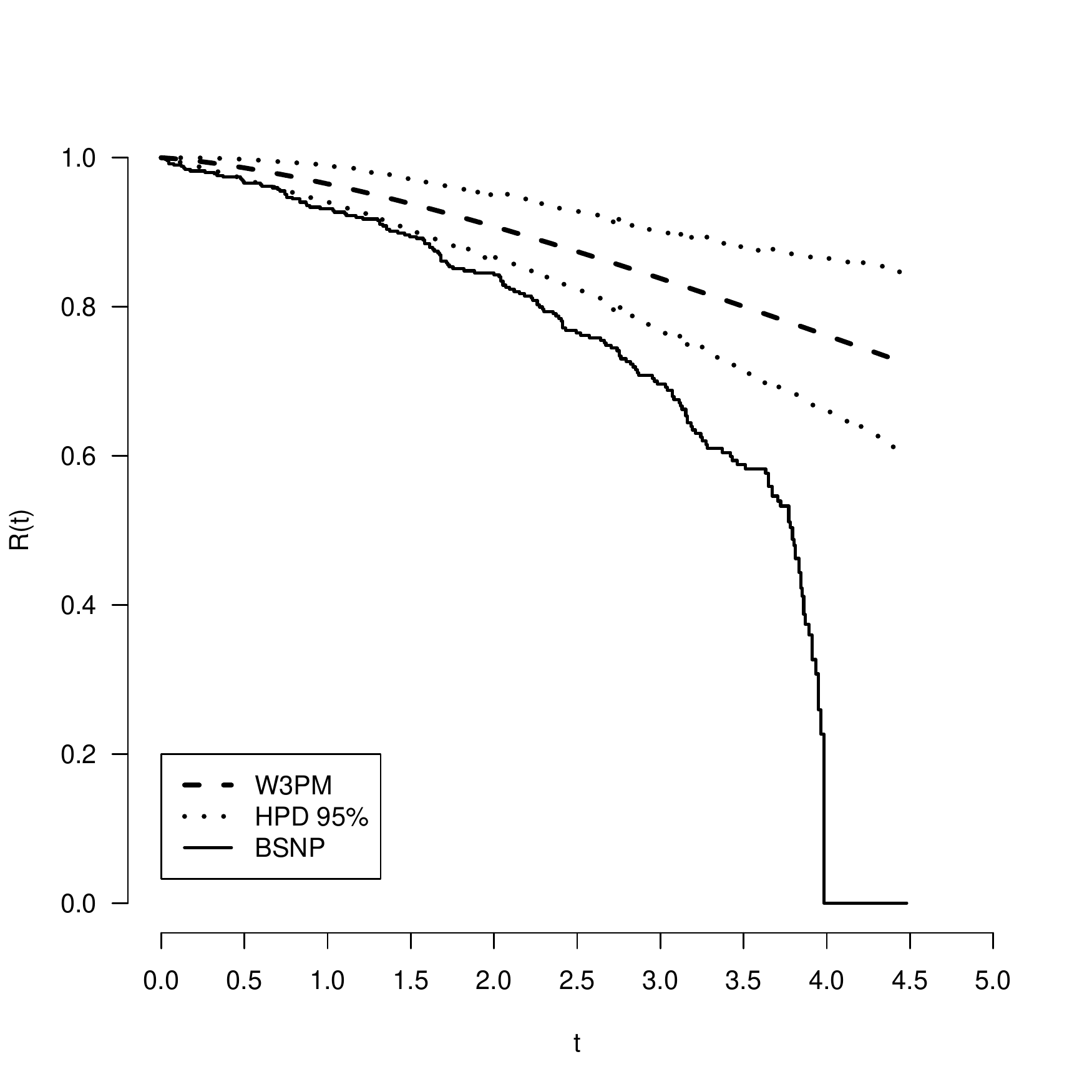}
		\subcaption{Head flyability ($j=2$)}
	\end{minipage}
	\begin{minipage}[b]{0.3\linewidth}
	  \includegraphics[width=\linewidth]{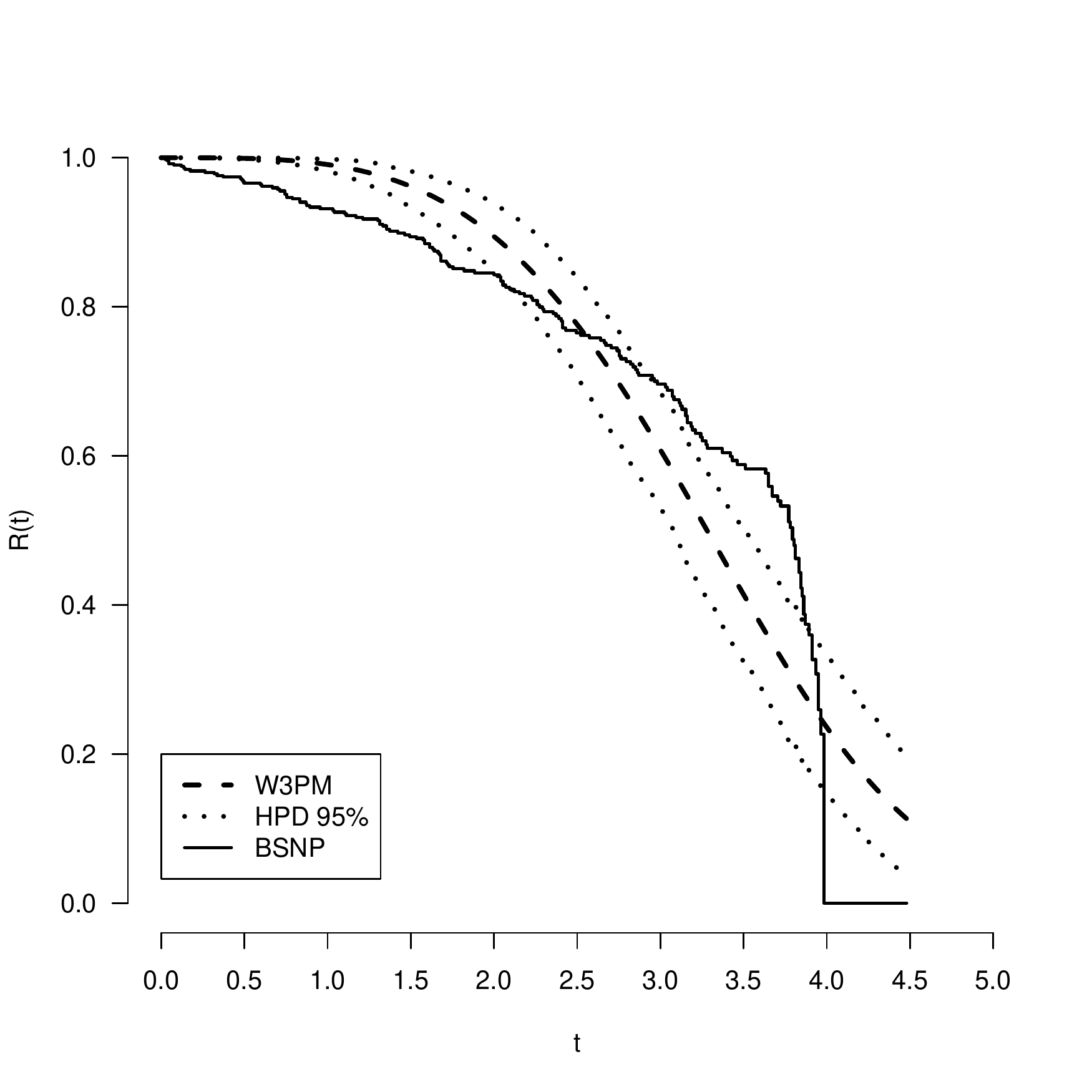}
      \subcaption{Head/disc magnetics ($j=3$)}
   \end{minipage}
	\caption{Estimated reliability curves by the proposed model and BSNP for three components involved in computer hard-drive.}
	\label{aplic_hard}
\end{figure}

In Figure \ref{aplic_hard} we present the estimated curves of eletronic hard, head flyability and head/disc magnetics by BSNP and proposed model. The proposed model under the symmetric assumption and not obtain overlapping curves and because of this only one of them is presented. As commented previously, BSNP estimates is the same for three failure cases and the proposed estimator is not close to the BSNP.  

\begin{table}[htbp]
	\centering
	\caption{Distribution of $n=172$ systems among causes of failure.}
	\begin{tabular}{cccccc}
		\hline
		& Failure by $j=1$   & Failure by $j=2$   & Failure by $j=3$   & $\Upsilon=\{1,3\}$ & $\Upsilon=\{1,2,3\}$ \\
		n (\%)  & 35 (20.35) & 19 (11.05) & 52 (30.23) & 32 (18.60) & 34 (19.77) \\
		\hline
	\end{tabular}%
	\label{casa_falha_drive}%
\end{table}%

\begin{table}[htbp]
	\centering
	\caption{Parameters posterior quantities for hard-drives application under symmetric assumption and not.}
	\begin{tabular}{ccccc}
		\hline
		\multicolumn{5}{c}{Component 1} \\
		\hline
		& \multicolumn{2}{c}{Symmetric Assumption } & \multicolumn{2}{c}{Symmetric Assumption Relaxation}  \\
		& Posterior mean & Posterior SD & Posterior mean & Posterior SD \\
		$\beta$ & 1.031 & 0.161 & 0.949 & 0.148 \\
		$\eta$ & 9.656 & 3.007 & 12.530 & 4.348 \\
		$\mu$   & 3.33E-38 & 6.11E-37 & 2.22E-47 & 2.95E-46 \\
		\hline
		\multicolumn{5}{c}{Component 2}  \\
		\hline
		& \multicolumn{2}{c}{Symmetric Assumption } & \multicolumn{2}{c}{Symmetric Assumption Relaxation}  \\
		& Posterior mean & Posterior SD & Posterior mean & Posterior SD \\
		$\beta$ & 1.531 & 0.309 & 1.490 & 0.314 \\
		$\eta$ & 10.490 & 4.455 & 9.595 & 3.480 \\
		$\mu$   & 1.29E-38 & 2.35E-37 & 2.17E-42 & 2.29E-41  \\
		\hline
		\multicolumn{5}{c}{Component 3}  \\
		\hline
		& \multicolumn{2}{c}{Symmetric Assumption } & \multicolumn{2}{c}{Symmetric Assumption Relaxation}  \\
		& Posterior mean & Posterior SD & Posterior mean & Posterior SD \\
		$\beta$ & 3.728 & 0.389 & 3.339 & 0.401 \\
		$\eta$ & 3.629 & 0.132 & 3.965 & 0.198 \\
		$\mu$    & 2.43E-08 & 1.46E-07 & 8.08E-13 & 3.12E-12  \\
		\hline
	\end{tabular}%
	\label{post_hard_drive}%
\end{table}%

\newpage
\section{Final Remarks} \label{final}
%
Before this work, no solution for more complex system (as bridge system) in masked data scenario was developed. With this motivation, a Bayesian three-parameter Weibull model for component reliability was proposed. The assumption of identical distributions of component lifetimes is not imposed. The presented model is said to be general because can be used for any coherent system, the symmetry assumption is not necessary and its application in acelerated life tests can be considered.  We worked with Weibull model; however it is quite simple to extend the work to other distributions or even to the pure likelihood approach.

The proposed model was compared to the nonparametric estimator proposed by \cite{Sassa} (BSNP) that can be considered for components involved in any system which the only necessary information is system failure time and structure. However, they assumed a restrictive assumption that components' lifetimes are {\it s}-independent and identically distributed. Because of this, there is only one estimator for all different components in the system. The simulation study had shown excellent performance of the proposed estimator and its superiority when compared to BSNP. The advantage of the proposed model is more evident as long as sample size increases. 

The practical relevance and applicability of the proposed model was demonstrated in a real dataset of computer hard-drives with three components in series. 

In this sense, the proposed estimator for component reliability function had demonstrated great performance in situations that lifetime distribution is not the same for all components in a coherent system with diferent proportion of masked systems. In estimation processes, satisfactory results about convergence were obtained and posterior quantities of reliability functions are easily obtained.

%
%

\newpage

\bibliographystyle{natbib}

\end{document}